\documentclass[paper,letterpaper]{JHEP}
\usepackage{epsfig,amssymb}

\newcommand{\be}{\begin{equation}}
\newcommand{\ee}{\end{equation}}
\newcommand{\bea}{\begin{eqnarray}}
\newcommand{\eea}{\end{eqnarray}}
\newcommand{\ie}{{\it i.e.}}
\newcommand{\eg}{{\it e.g.}}
\newcommand{\Z}{\mathbb{Z}}
\newcommand{\R}{\mathbb{R}}
\newcommand{\C}{\mathbb{C}}
\newcommand{\U}{\mathop{\rm U}}
\newcommand{\SU}{\mathop{\rm SU}}
\newcommand{\SO}{\mathop{\rm SO}}

\newcommand{\Sp}{\mathop{\rm Sp}}
\newcommand{\Rea}{\mathop{\rm Re}}
\newcommand{\Ima}{\mathop{\rm Im}}
\newcommand{\Nf}{${\cal N}{=}4$}
\newcommand{\Nt}{${\cal N}{=}2$}
\newcommand{\del}{\partial}
\newcommand{\vn}{{\vec\nabla}}
\newcommand{\vB}{{\vec B}}
\newcommand{\vE}{{\vec E}}
\newcommand{\vF}{{\vec F}}
\newcommand{\vx}{{\vec x}}
\newcommand{\bee}{{\bf e}}
\newcommand{\bX}{{\bf X}}
\newcommand{\bxi}{\mbox{\boldmath$\xi$}}

\title{String Webs from Field Theory}
\author{Philip C. Argyres and K. Narayan\\
        Newman Laboratory, Cornell University, 
        Ithaca NY 14853\\[1mm]
        E-mail: \email{argyres@mail.lns.cornell.edu,
	narayan@mail.lns.cornell.edu}}

\abstract{The spectrum of stable electrically and magnetically charged
supersymmetric particles can change discontinuously as one changes the
vacuum on the Coulomb branch of gauge theories with extended
supersymmetry in four dimensions.  We show that this decay process can
be understood and is well described by semiclassical field
configurations purely in terms of the low energy effective action on
the Coulomb branch even when it occurs at strong coupling.  The
resulting picture of the stable supersymmetric spectrum is a
generalization of the ``string web'' picture of these states found in
string constructions for certain theories.}

\preprint{CLNS-00/1704}

\begin{document}   

\section{Introduction}\label{s1}

Four dimensional gauge theories with at least eight supersymmetries
have a Coulomb branch of inequivalent vacua in which the low energy
effective theory generically has an unbroken $\U(1)^n$ gauge
invariance.  These theories also have a spectrum of massive charged
particles with various electric and magnetic charges under the
$\U(1)$'s, lying in supersymmetry multiplets.  Those lying in short
multiplets of the supersymmetry algebra (the BPS states) leave some
fraction of the supercharges unbroken, and their masses are related to
their charges by the supersymmetry algebra \cite{wo78}.  The spectrum
of the possible BPS masses can then be determined using supersymmetry
selections rules \cite{sw9407,sw9408}.  This, however, leaves open the
question of the existence and multiplicity of these states.
Furthermore, even if such a state exists in some region of the Coulomb
branch (for some values of the vevs), they may be unstable to decay at
curves of marginal stability (CMS) on the Coulomb branch
\cite{cv9211,sw9407}.  In this paper we propose a solution to the
question of the multiplicity of BPS states for \Nt\ and 4 theories in
four dimensions just in terms of the low energy effective $\U(1)^n$
action on the Coulomb branch.

The form of the answer we get coincides with the ``string web''
picture of BPS states
\cite{f9701,b9712,ghz9801,bf9802,mns9803,bk9804,dhiz9805,bf9806,bhlms9901,t9902}
developed in the context of the D3-brane construction of \Nf\ $\SU(n)$
superYang-Mills (SYM) theory \cite{w9510}, and the F theory solution
to \Nt\ $\SU(2)$ gauge theory with fundamental matter
\cite{s9605,bds9605}.  (In general, it is possible to realize certain
classes of gauge theories as worldvolume field theories on probe
D3-branes in specific string theory backgrounds in the limit of
vanishing string length, $\ell_s \to 0$---see \cite{gk9802} for a
review of these constructions.)  BPS states carrying electric and
magnetic charges $(p,q)$ in the low energy theory (which is the
effective action on the brane probe) are realized in these
constructions as webs of $(p,q)$ strings meeting at 3-string junctions
and ending on the probe D3-brane as well as various ``sources'' in the
string theory background space time.

\EPSFIGURE{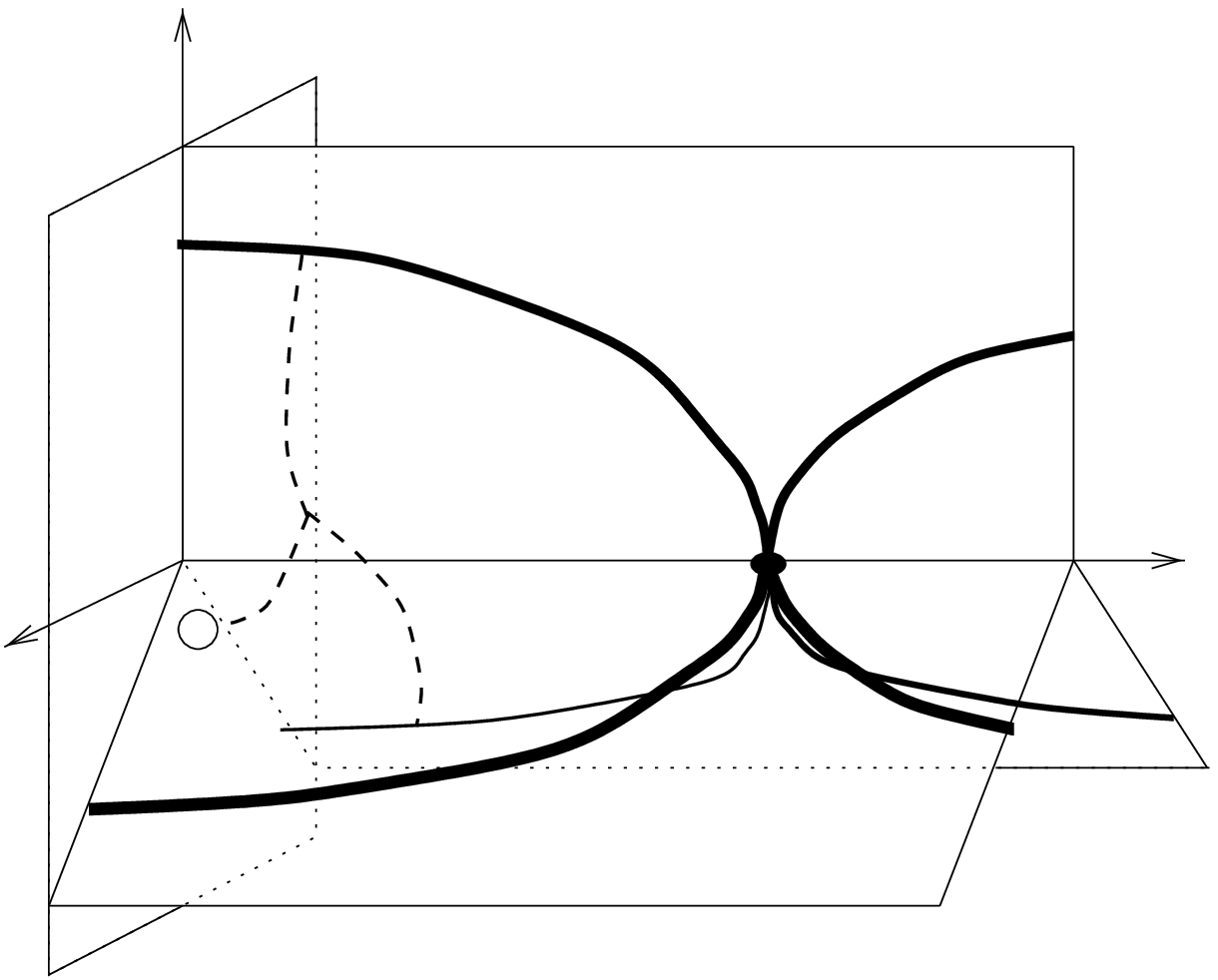,width=16em}{A slice of the Coulomb branch of
$SU(3)$ \Nt\ SYM showing some complex co-dimension one curves of
singularities (dark lines), and a BPS state represented as a string
web (dashed lines) joining a given vacuum (the open
circle).\label{fig1}}

Our solution coincides with the string web constructions in the cases
mentioned above, and generalizes these constructions to arbitrary
field theory data (gauge groups, matter representations, couplings and
masses).  The resulting picture is quite simple: BPS states are
represented by string webs on the Coulomb branch of the theory with
one end at the point corresponding to the vacuum in question (the
analog of the 3-brane probe in the F theory picture) and the other
ends lying on the complex codimension 1 singularities on the Coulomb
branch (the analogs of the $(p,q)$ 7-branes of the F theory picture).
The strands of the string web lie along geodesics in the Coulomb branch
metric.  Each strand carries electric and magnetic charges
under the $\U(1)^n$ low energy gauge group: the total charge flowing
into the vacuum point determines the total charge of the BPS state,
while only multiples of the charges determined by the $\Sp(2n,\Z)$
monodromies around the codimension 1 singularities are allowed to flow
into those ends of the web; see figure~\ref{fig1}.  Finally, three string
junctions obey a tension-balancing constraint, where the tension of
the strings is given by the usual formula in terms of the electric
and magnetic charges they carry.

Perhaps the most surprising thing about our solution is that it
describes the stability of the monopole and dyon BPS spectrum wholly
in terms of the $\U(1)^n$ low energy effective action.\footnote{Note,
however, that the original four dimensional studies \cite{fb9602} of
the BPS stability question already used the low energy effective
action---in particular the global structure of the Coulomb branch---to
determine the stable BPS spectrum in $\SU(2)$ gauge theories.}  This
is possible because the distance $\Delta X$ from a given vacuum on the
Coulomb branch to a CMS acts as a new low energy scale which can be
made arbitrarily small compared to the strong coupling scale $\Lambda$
of the nonabelian gauge theory as we approach the CMS.  In particular,
we will show that, as we approach the CMS, the classical low energy
field configuration describing a state which decays across the CMS
will develop two or more widely separated charge centers (which appear
as singularities in the low energy solution); the distance between
these centers varies inversely with $\Delta X$.  Thus, in this limit
the details of the microscopic physics become irrelevant for the decay
of a BPS state across a CMS, which is described by a low energy field
configuration with charge centers becoming infinitely separated.

In general, classical BPS field configurations in the low energy
effective field theory can be interpreted as a static deformation of a
3-brane worldvolume known as a brane spike \cite{cm9708}, or as a
``brane prong'' in the case where the field configuration has multiple
sources.  In the limit that the cutoff scale is removed ({\em e.g.},
$\Lambda \to \infty$ or $\ell_s \to 0$) such a brane spike or prong
approaches a $(p,q)$ string or string web ending on the 3-brane.  Thus, in
the limit that we approach the CMS this picture becomes arbitrarily
accurate in the $\U(1)^n$ effective theory; it can then be extended to
the rest of the Coulomb branch essentially by analyticity.

The brane prong picture arises from thinking of the classical BPS
field configuration of the scalar fields in the low energy
supersymmetric $\U(1)^n$ gauge theory on the Coulomb branch as maps
from space-time (the worldvolume of the 3-brane) to the Coulomb branch
(the background geometry that the 3-branes live in).  Such a field
configuration will have one or more singularities or sources where
scalar field gradients and $\U(1)$ field strengths diverge.  Near
these points the low energy description breaks down and should be cut
off by boundary conditions reflecting the matching onto the
microscopic physics of the nonabelian gauge theory.  We will show that
these boundary conditions are essentially determined by the BPS
condition: all other details of the choice of boundary conditions do
not affect the behavior of the prong solution in the limit that the
cut off length scale vanishes relative to the low energy length scale.
In particular, since the ratio of the cut off length scale to the
relevant low energy length scale (the distance between the separating
sources) vanishes at the CMS, the way the BPS spectrum jumps across
the CMS is independent of the the details of the way the BPS charges
are regularized.

This paper is organized as follows.  In the next section we present a
simple physical argument for why BPS states near a CMS develop widely
separated charge centers and therefore have a good semiclassical
description in the low energy theory.  In section~\ref{s3} we solve
for the semiclassical field configurations and show the decay of the
relevant BPS states in a simple $\U(1)$ toy model.  In particular, the
separation of the charge centers near the CMS can be simply and
explicitly demonstrated in this example.  In section~\ref{s4} we
generalize our analysis to the $\U(1)^n$ low energy theory of \Nf\ SYM,
and show how the string web picture of $1/2$ and $1/4$ BPS states is
recovered in the $\SU(N)$ case. 
Section~\ref{s5} discusses the generalization to \Nt\ theories.
Section~\ref{s6} closes with some open questions and directions for
future research: the most pressing open question has to do with the
derivation of the ``s-rule'' \cite{hw9611,mns9803,dhiz9805,bf9806} in \Nt\
theories, which is not apparent in our solutions; interesting
extensions of the arguments of this paper apply to similar phenomena
in other dimensions and in gravitational theories.

The work in this paper overlaps that of a number of other papers which
have appeared in the last few years.  The main new contribution of
this paper is the understanding that a string web picture of decaying
BPS states follows directly from the low energy effective theory in
the vicinity of a CMS and relies on approximate boundary conditions
which become exact in the limit of approaching the CMS.  More
specifically, the semiclassical description of BPS states in low
energy $\U(1)^n$ effective theories has been explored in many
contexts, especially \cite{gkmtz9903} whose discussion of brane prong
solutions was a starting point for this paper, and \cite{kk0002} whose
discussion of how brane prong solutions approximate string webs near
the CMS overlaps with our discussion in section~\ref{s3}.  A related
discussion of brane spikes in an \Nt\ F theory background appears in
\cite{kpr9903}.  Our discussion of brane spike and prong solutions
differs mainly in our treatment of the boundary conditions at the
charge sources as well as in our generalization of these solutions to
arbitrary \Nf\ or \Nt\ supersymmetric field theory data.  Also, the
basic phenomenon of separating charge centers (or at least the growth
in overall size) of BPS states near CMS has been noted repeatedly in
the context of semiclassical dyon solutions in SYM theories
\cite{hhs9803,ko9804,ly9804,blly9906,bly9907,bl9909,gkpy9912,gkly0008}.
In particular the general picture of loosely bound composite BPS
states in \cite{rsvv0006} overlaps with our discussion in
section~\ref{s2}; we add to it the observation that the separation of
the charge centers is generic and persists in strong coupling regions
of the Coulomb branch.  Finally, the discussion in the last section of
\cite{d0005} amounts to a gravitational version of our discussion in
section~\ref{s5}; in addition to our different treatment of the
boundary conditions following from the regime of validity of the low
energy effective theory, we add the observation that the image of the
low energy solution in moduli space more and more closely approximates
a string web configuration as we approach the CMS.

\section{BPS states near CMS}\label{s2}

Since the total electric and magnetic charges $(Q_E^i,Q_{Bi})$ with
respect to the unbroken $U(1)^n$ gauge group on the Coulomb branch are
conserved, we can restrict our attention to a single charge sector of
the theory.  Our question is whether at a given vacuum there is or is
not a one-particle BPS state in that charge sector.  The mass of a BPS
state is determined by the superalgebra to be the absolute value of
the central charge, and the central charge is the sum of terms
proportional to the charges
\be
Z = Q_E^i a_i + Q_{Bi} a_D^i
\ee
where the coefficients $a_i$ and $a_D^i$ depend only on the vacuum in
question and not on the charges.  This mass, $|Z|$, is the minimum
mass of any state (BPS or not, single particle or not) in this charge
sector and so, in particular, the spectrum in this charge sector is
gapped.

\EPSFIGURE{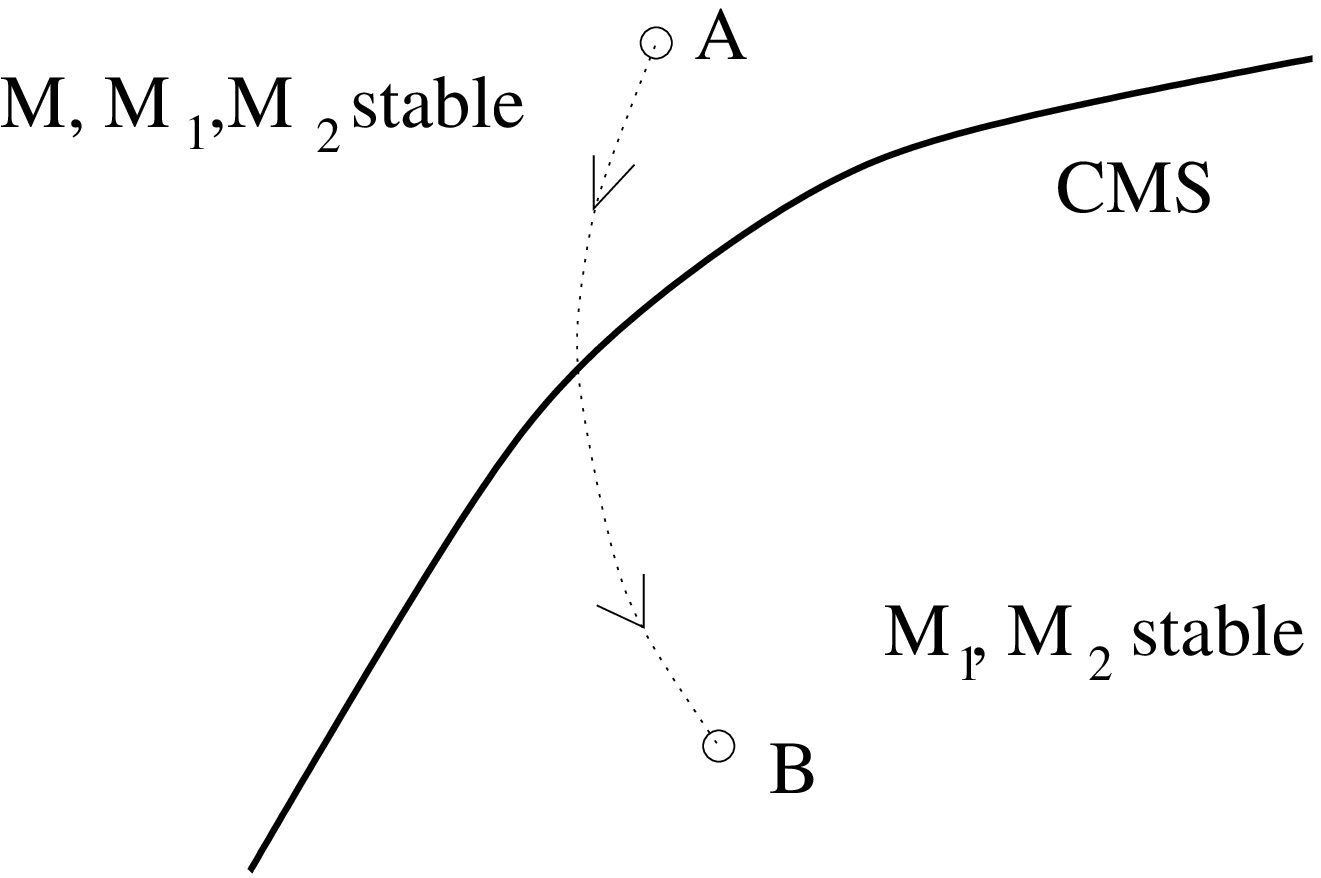,width=16em}{A generic decay across a CMS.
The dotted line denotes a path on the Coulomb branch.\label{fig2}}

A single BPS particle in this charge sector would have a mass $M=|Z|$.
It is stable or at worst marginally stable against decay into two (or
any number of) constituent particles, since by charge conservation $Z =
Z_1 + Z_2$, so by the triangle inequality
\be\label{trineq}
M \le M_1 + M_2.  
\ee
The CMS are submanifolds of the Coulomb branch where the inequality in
(\ref{trineq}) is saturated.  As one adiabatically changes the order
parameter on the Coulomb branch from a vacuum $A$ on one side of a CMS
to a vacuum $B$ on the other, the one particle state $M$ in our charge
sector will become more and more nearly degenerate with the two
particle state $M_1+M_2$.  Supposing $M$ does decay across the CMS,
then only the two particle state $M_1+M_2$ will remain in the spectrum
once we have crossed the CMS.  This is illustrated in
figure~\ref{fig2}.  By assumption $M_1$ and $M_2$ are in the stable
spectrum everywhere on the $B$ side of the CMS.  It follows from the
BPS mass formula that they are also generically stable on the $A$
side, since possible decays like $M_1 \to M + (-M_2)$ (where $-M_2$ is
the charge conjugate of $M_2$) are not allowed since $M_1 < M + M_2$
on the CMS.  Also, it follows from (\ref{trineq}) that the
two particle state $M_1 + M_2$ is not BPS, even for zero relative
momentum, except precisely at the CMS.

In terms of the density of states, on the $A$ side of the CMS where
$M$ is stable, we have a discrete one particle state lying below the
threshold to the $M_1+M_2$ two particle state continuum.  Right at the
CMS the one particle state just coincides with the two particle
threshold.  On the $B$ side of the CMS, where $M$ is no longer in the
spectrum, there is just the two particle continuum; see
figure~\ref{fig3}.   

Since the transition takes place right at threshold, $M$ will
``decay'' into the two particle state with zero relative momentum.
Now, it follows from (\ref{trineq}) that the two particle state $M_1 +
M_2$ is not BPS, even for zero relative momentum, except precisely at
the CMS.  Thus there will generically be no BPS force cancelation
between particles $M_1$ and $M_2$, so the zero relative momentum two
particle state is classically approximated by two spatially infinitely
separated one particle states (to have a static configuration).

\EPSFIGURE{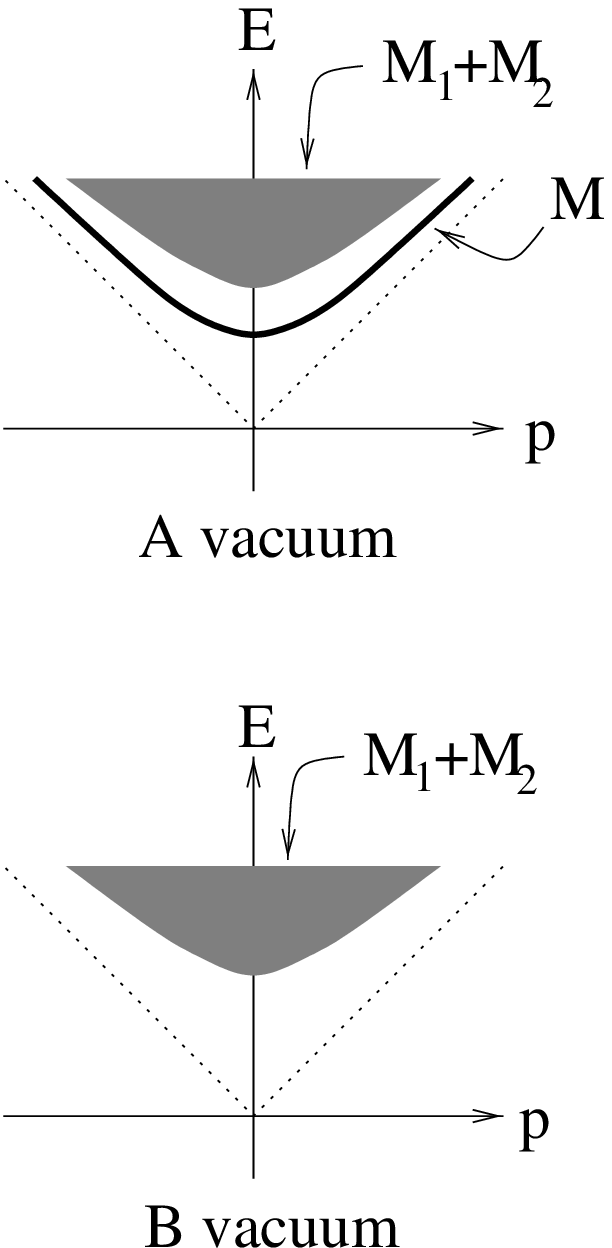,width=12em}{Generic density of states in
the charge sector of $M$ at the $A$ and $B$ vacua.\label{fig3}}

The transition across the CMS of a one particle state to a widely
separated, zero momentum, two particle state should go by way of field
configurations with large spatial overlap, as follows heuristically
from locality and the adiabatic theorem.  In particular, this
implies that a state decaying just where its mass reaches the
two-particle threshold of its decay products (zero ``phase space'')
should have a diverging spatial extent as it approaches the
transition.  This is despite the fact that precisely at the CMS the
two particle state is BPS and so may be spatially small: the relevant
field configurations are the ones with widely separated centers which
have large overlap with the two particle decay state away from the
CMS.

The large spatial extent of a state decaying at a CMS means that the
decay of such a state should be visible semiclassically in the low
energy effective action.  We will show in a simple $\U(1)$ example in
the next section that this is, indeed, the case.  This argument gives
the basic physics underlying our approach to decays across CMS:
{\em if a BPS state does decay across a CMS, that decay will be visible
semiclassically in the low energy effective action even if it
takes place at strong coupling from a microscopic point of view.}

We can analyze this picture further to predict in more detail what we
should see in the classical BPS field configurations of the low energy
$\U(1)^n$ effective action on the Coulomb branch.  Since the two 
particle zero momentum state is classically given by two static charges
at very large spatial separation, its corresponding classical
low energy field configuration is just the long-range response of the
massless fields to the charge sources.  Thus we expect the decaying
one particle state to be a static dumbbell-like configuration of
the massless fields, \ie\ one with two charge centers whose relative
separation $|\vx_1-\vx_2| \sim 1/\Delta X$ diverges as we approach
the CMS: $\Delta X \to 0$.  Here $X$ represents the vevs of the scalar
fields parameterizing the Coulomb branch.

The charge centers themselves are singularities in the low energy
description.  They should be regularized at an appropriate microscopic
length scale, $1/\Lambda$, where $\Lambda$ is some strong coupling
scale of the underlying gauge theory.  This cut off scale and the
boundary conditions for the massless fields near the charge cores will
be discussed in detail in the next section.  Since the mass gap
between the two particle threshold and the one particle state
decreases as we approach the CMS, the two charge cores must be more
and more loosely bound in this limit.  Indeed, the low lying field
theory spectrum in the given charge sector should be approximated by
the spectrum of excitations around our static ``soliton''
configuration.  Quantizing the collective motions of the soliton will
reproduce the diminishing mass gap and the two particle continuum in
the limit as we approach the CMS: the extra three (non-relativistic)
degrees of freedom of the two particle continuum come from the two
rotational and one vibrational mode of the dumbbell configuration.  The
increasing separation of the charge (and mass) centers and the
weakening of their binding implies that the energy level spacing and
gap of this rotator and oscillator indeed vanishes at the
CMS.\footnote{This point is made more concretely (and
supersymmetrically) in \cite{rsvv0006} in semiclassical computations
at weak coupling.}

The resulting picture is quite intuitive: a BPS state decaying across
a CMS does so by becoming an ever larger, more loosely bound state of
its eventual decay products.  Once the CMS is crossed, the bound state
ceases to exist, and so, in particular, there will be no static BPS
solutions to the low energy equations of motion and boundary
conditions in this region of the Coulomb branch.

\section{A $\U(1)$ toy example}\label{s3}

In this section we will consider the low energy $\U(1)$ effective
action with two real scalars $X_r$, $r=1,2$:
\be\label{toym}
S = -\int d^4 x \left( {1\over4} F_{\mu\nu} F^{\mu\nu}
+ {1\over2} \del_\mu X_1 \del^\mu X_1 + {1\over2} \del_\mu X_2 \del^\mu X_2 \right)
\ee
This can be thought of either as the bosonic sector of a $\U(1)$ \Nf\
SYM theory with four of the six (real) scalars suppressed, or as that
of a $\U(1)$ \Nt\ SYM theory with a flat Coulomb branch.  The gauge
coupling has been set to unity for simplicity and a theta angle term
does not affect the energy functional which is really all we will be
working with, so it has not been included in this expression.  We will
be looking for static solutions to this theory, and will denote
spatial vectors by arrows, \eg\ $\vx$, and vectors on the $X_1$--$X_2$
plane by boldface letters, \eg\ $\bX$.  The static sourceless
equations of motion are $\vn\cdot\vE =\vn\times\vE =\vn\cdot\vB
=\vn\times\vB =0$ and $\nabla^2 \bX = 0$, where the electric $\vE$ and
magnetic $\vB$ fields have been defined in the usual way.  If a region
contains a charge core with electric and magnetic charges $(Q_E,Q_B)$,
then by Gauss' law we have
\be\label{gauss} 
\oint_{S^2_\Lambda} \vE\cdot d{\vec a} = Q_E, \qquad 
\oint_{S^2_\Lambda} \vB\cdot d{\vec a} = Q_B, 
\ee 
where the integral is over a sphere $S^2_\Lambda$ of radius $r_\Lambda$
enclosing the charged core.

The toy theory (\ref{toym}) by itself is too simple to be interesting,
but all we need to add to it to capture the essential physics of the
decay of BPS states across CMS are the presence of singularities in
the vacuum manifold.  In this case the vacuum manifold (Coulomb
branch) is the $\bX$ plane.  In actual \Nt\ or \Nf\ SYM theories
there are complex submanifolds of the Coulomb branch which are
singularities in the effective action since they correspond to vacua
at which charged states become massless, and so must be included in
the effective action.  In \Nt\ theories these appear as curvature
singularities in the low energy $\U(1)^n$ effective action sigma model
metric on the Coulomb branch; in \Nf\ theories the Coulomb branch is
flat and the singularities appear as orbifold fixed points (\ie\
curvature delta functions).  For convenience and concreteness, we will
include such singularities in our toy model (\ref{toym}) by simply
positing that there are two singularities at points on the Coulomb
branch with coordinates
\be\label{toysing1}
\bX=\bX^E\equiv(L,0)
\ee
where a particle with electric charge
\be\label{toysing2}
(Q_E,Q_B)=(1,0)
\ee
becomes massless, and
\be\label{toysing3}
\bX=\bX^B\equiv(0,L)
\ee
where a particle with magnetic charge
\be\label{toysing4}
(Q_E,Q_B)=(0,1)
\ee
becomes massless.

In section~\ref{s4} we generalize the calculations of this section to the
full $\U(1)^n$ \Nf\ effective action with $6n$ free real scalar
fields, and in section~\ref{s5} to the $\U(1)^n$ \Nt\ effective action
with $2n$ real scalar fields with curved sigma model metric.  In both
cases the core of the physics will be seen to be the same as that of
the toy model of eqns.~(\ref{toym}) and
(\ref{toysing1}--\ref{toysing4}), though the combinatorics and some
details of the analysis will be more complicated.

\subsection{Boundary conditions and BPS bounds}

As a low energy effective action, (\ref{toym}) should be considered as
the first terms in a derivative expansion.  Suppressing all the higher
derivative terms is a good approximation as long as we do not probe
the physics describing the core of the BPS state.
Near such a core the $\U(1)$ field strength and, by supersymmetry,
the values of the scalar fields $\bX$ will grow large.  Indeed,
in the presence of a static source at spatial position $\vx_0$
the electric and magnetic fields and the scalar fields diverge as
\be
|\vE|, \ |\vB|,\ |\vn \bX| \ \sim \ {1\over r^2},
\ee
where $r = |\vx - \vx_0|$.  With an ultraviolet scale $\Lambda$ 
suppressing derivative corrections to the $\U(1)$ effective 
action by powers of $\Lambda$, for example $\Lambda^{-2} |\del \bX|^4$,
we see that when $|\vn \bX| \sim \Lambda^2$, or at a typical distance
\be\label{rla}
r \sim r_\Lambda \equiv {1\over\Lambda}
\ee
{}from the core, the low energy $\U(1)$ solution ceases to be valid.
Also, when expanding about a given vacuum $\bX^0$ on the Coulomb
branch there will also be irrelevant terms with polynomial
coefficients, \eg\ $\Lambda^{-6} |\bX-\bX^0|^2 |\del \bX|^4$, implying the
breakdown of the low energy solution when $|\bX-\bX^0| \gtrsim \Lambda^3
r^2$ where $r$ is the spatial distance from a charge core.
Such terms imply that the value $\bX(\vx)$ of the
scalars at any given spatial position $\vx$ is itself reliable only
up to some accuracy
\be
|\delta \bX(\vx)| \simeq \widehat\Lambda .
\ee
This fuzziness in the solution $\widehat\Lambda$ is not easy to estimate
directly from the action, for it varies with $\vx$ as well as with
the choice of vacuum $\bX^0$ and charges $(Q_E,Q_B)$.  We will see
below how it can be determined self-consistently from static solutions.

Our strategy will be to use the low energy $\U(1)$ solution away from
the cores ($r>r_\Lambda$) and impose appropriate boundary conditions in
the vicinity of the charge cores ($r\sim r_\Lambda$).  The qualitative
features of these boundary conditions are easy to deduce, as we will
discuss momentarily; we will see in the course of this section that
only these qualitative features are important for the physics in the
vicinity of a CMS---other details of the boundary conditions do not
affect the results.

The basic boundary condition (\ref{gauss}) for the $\U(1)$ gauge field
follows from charge conservation.  The boundary conditions for the
scalar fields $\bX$ are more subtle.  Suppose we are looking for a
solution with only electric charge $(Q_E,Q_B)=(1,0)$.  The solutions
to the equation of motion near a point-like charge source together
with supersymmetry imply that $|\bX|$ diverges as one approaches the
source.  On the cutoff sphere $S^2_\Lambda$ around the source $\bX$ takes
some finite values.  Suppose they are approximately constant, $\bX =
\bX_\Lambda$ on $S^2_\Lambda$.  The solution interior to $S^2_\Lambda$ is then
approximately that of a state with charge $(1,0)$ in the vacuum
$\bX_\Lambda$.  The minimum mass of this state is given by its BPS mass
$M_{(1,0)}$ which is some function of the vacuum $\bX_\Lambda$.  This mass
is minimized at the singularities $\bX = \bX^E$ where the $(1,0)$
charged BPS states become massless by assumption.  Thus energy
minimization implies that the scalar fields satisfy the approximate
Dirichlet boundary condition
\be\label{dirbcE} 
\bX \simeq \bX^E\ \ \mbox{within}\ \ 
B^3_\Lambda\ \ \mbox{for electric charges},  
\ee
and a similar argument implies
\be\label{dirbcB} 
\bX \simeq \bX^B\ \ \mbox{within}\ \ 
B^3_\Lambda\ \ \mbox{for magnetic charges},
\ee
where $B^3_\Lambda$ is a ball of approximate radius $r_\Lambda$ around each
charge core and $\bX\simeq\bX_E$ means only that $\bX$ pass within
distance $\widehat\Lambda$ to $\bX_E$ on the Coulomb branch.  We will call
these conditions on the scalar fields ``fuzzy ball'' boundary
conditions, a kind of weak form of Dirichlet boundary conditions.
Figure~\ref{fig5} below gives an illustration of these fuzzy
ball boundary conditions on the Coulomb branch in our toy example.

So far we have only argued that the fuzzy ball boundary conditions
express a {\em tendency} for the scalar field to approach those
values.  The key step in our analysis of BPS states near CMS is to
{\em assume} the above fuzzy ball boundary conditions on the scalar
fields.  The idea is that the fuzzy ball boundary condition is all we
can physically demand of the low energy solution since it is not
accurate on spatial resolutions less than $r_\Lambda$ nor for field value
resolutions less than $\widehat\Lambda$.  We then use them to solve for BPS
configurations of charge $(Q_E,Q_B)$ by positing some number of
electric sources of total charge $Q_E$ and magnetic sources of total
charge $Q_B$, minimizing their energies, and checking that the fuzzy
ball boundary conditions are self consistent, \ie\ that for a spatial
cutoff length scale $r_\Lambda$ consistent with (\ref{rla}) the static
solution has $\widehat\Lambda \ll |\bX^E-\bX^B|$.  

In the limit as the vacuum approaches a CMS, as we have described
qualitatively in the last section and will see explicitly below, the
size of the field configuration grows, and so the relative size of the
cutoff region $r_\Lambda$ to the field configuration shrinks.  Thus in
this limit one expects that our fuzzy ball boundary condition to
become a simple Dirichlet boundary condition, \ie\ $\widehat\Lambda\to0$.
Indeed, as shown in \cite{gkmtz9903} for spherically symmetric 1/2 BPS
states in \Nf\ SYM theory, a simple (spherical) Dirichlet boundary
condition exactly reproduces the BPS bound; near the CMS a decaying
1/4 BPS state will become to arbitrary accuracy a very loosely bound
state of two 1/2 BPS states, and so the fuzzy ball boundary conditions
should approach Dirichlet boundary conditions to the same accuracy.

Furthermore, far enough away from the CMS (at distances greater than
or on the order of $\Lambda$ on the Coulomb branch) the boundary
conditions (\ref{dirbcE}) and (\ref{dirbcB}) will break down entirely,
and cannot be satisfied even approximately.  Thus, far out on the
Coulomb branch, where monopole states are well-described by
semiclassical nonabelian field configurations, our low energy $\U(1)$
description breaks down entirely.\footnote{More precisely, it becomes
equivalent to the Dirac monopole solution which carries no
information about the structure of the state.}  In this sense our
description of BPS states is complementary to the semiclassical one.

Now let us use these boundary conditions to derive the low energy
field equations satisfied by our static BPS solutions.  We do this by
the familiar method of minimizing the energy of the field
configurations.  From the action (\ref{toym}), an energy density
functional for static configurations can be constructed by the usual
canonical methods
\be
{\cal E} = {1\over2} \left[ \vE^2 + \vB^2 + (\vn \bX)^2 \right] .
\ee
This can be rewritten in the following fashion
\bea
{\cal E} &=& {1\over2} \left[ (\vE - \cos \alpha \vn X_1 + \sin\alpha 
\vn X_2)^2 + (\vB - \sin \alpha \vn X_1 - \cos \alpha \vn X_2)^2 \right]
\nonumber \\
&& \quad {}+ \cos \alpha (\vE \cdot \vn X_1 + \vB \cdot \vn X_2) 
+ \sin \alpha (\vB \cdot \vn X_1 - \vE \cdot \vn X_2).
\eea
Integrating this expression over three dimensional space
gives the mass of the configuration as
\bea
M &=& \int d^3\vx\, {1\over2} \left[ (\vE - \cos \alpha \vn X_1 + \sin\alpha 
\vn X_2)^2 + (\vB - \sin \alpha \vn X_1 - \cos \alpha \vn X_2)^2 \right]
\nonumber \\
&& \qquad {}+ \sum_{I=0}^n \left\{
\cos\alpha \oint_{S^2_I} (X_1\vE+X_2\vB)\cdot d{\vec a} + 
\sin\alpha\oint_{S^2_I} (X_1\vB-X_2\vE)\cdot d{\vec a}\right\}
\eea
where the boundaries $S^2_I$ are spheres around each charge source and
one sphere at infinity.  We have used the divergence-free equation of
motion for the electric and magnetic fields $\vn\cdot\vE =\vn\cdot\vB
=0$ away from the sources.  If we label the boundaries so that the
$I=0$ boundary is the one at infinity and the $I=i\neq0$ are the ones
around the charge sources, then our approximate Dirichlet boundary
conditions (\ref{dirbcE}) and (\ref{dirbcB}) imply that at the $i$th
boundary the scalars $\bX$ take the constant values $\bX^i=\bX^E$ or
$\bX^B$ while at infinity they take their asymptotic values $\bX^0$,
the Coulomb branch coordinates of the vacuum.  Since these are
constants they can be taken outside the surface integrals which then
give by (\ref{gauss}) the charges $(Q^I_E, Q^I_B)$ enclosed by each
sphere, so that
\bea\label{bpsmass}
M &=& \int d^3\vx\, {1\over2} \left[ (\vE - \cos \alpha \vn X_1 + \sin\alpha 
\vn X_2)^2 + (\vB - \sin \alpha \vn X_1 - \cos \alpha \vn X_2)^2 \right]
\nonumber \\
&& {}+ \sum_{I=0}^n \left[\cos\alpha(X_1^I Q_E^I+X_2^I Q_B^I) + 
\sin\alpha(X_1^I Q_B^I-X_2^I Q_E^I)\right] .
\eea
Here $(Q_E^0,Q_B^0) = (Q_E, Q_B)$ is the total charge of the
configuration.  By charge conservation
\be\label{chrcons}
(Q_E,Q_B) = -\sum_{i=1}^n (Q_E^i,Q_B^i)
\ee
so, defining the position vectors of the singularities on the
Coulomb branch relative to the vacuum by
\be
\bxi^i \equiv \bX^i-\bX^0,
\ee
and since the first line in (\ref{bpsmass}) is positive definite
we get the bound
\be
M \ge \cos\alpha(\xi_1^i Q_E^i + \xi_2^i Q_B^i)
+ \sin\alpha(\xi_1^i Q_B^i - \xi_2^i Q_E^i),
\ee
where the sum on $i$ over the charge sources is implied.  The
tightest bound on $M$ comes from maximizing the right hand side
with respect to $\alpha$, giving the BPS bound
\be \label{M}
M \ge \sqrt{(\xi_1^i Q_E^i + \xi_2^i Q_B^i)^2 
+ (\xi_1^i Q_B^i - \xi_2^i Q_E^i)^2} .
\ee
Thus the minimal energy configurations saturating the 
inequality are solutions of the BPS equations
\bea\label{EB}
\vE &=& \cos \alpha \vn X_1 - \sin \alpha \vn X_2, \nonumber \\
\vB &=& \sin \alpha \vn X_1 + \cos \alpha \vn X_2
\eea
where
\be\label{tanal}
\tan \alpha = {\xi_1^i Q_B^i - \xi_2^i Q_E^i \over
\xi_1^i Q_E^i + \xi_2^i Q_B^i}.
\ee
The rest of this section is devoted to analyzing the solutions to
(\ref{EB}) subject to our boundary conditions (\ref{dirbcE}) and
(\ref{dirbcB}) minimizing the BPS mass (\ref{M}).

\EPSFIGURE{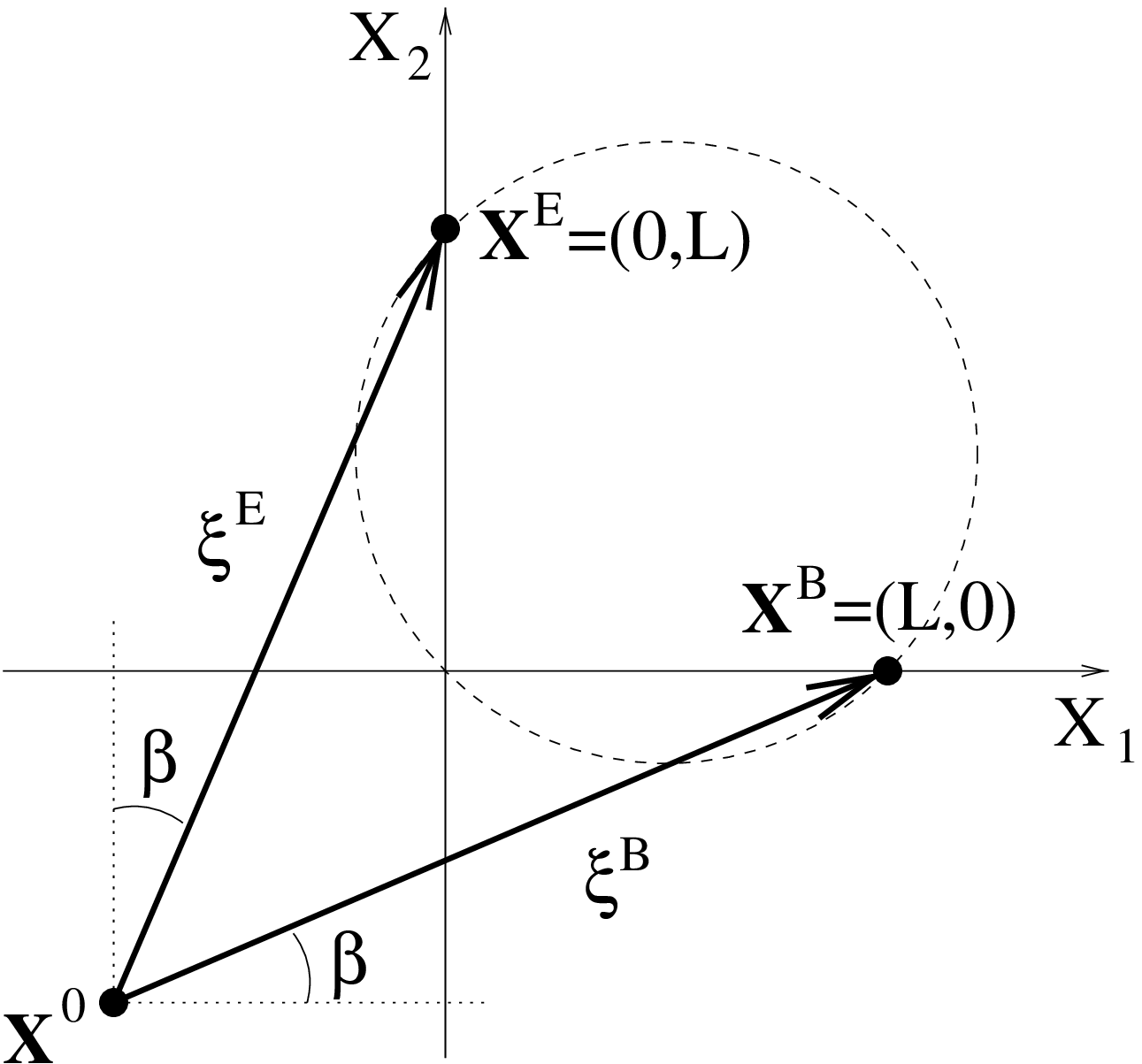,width=16em}{Coordinates on the Coulomb branch
of the $\U(1)$ toy model.  $\bX=\bX^0$ denotes the vacuum, while
$\bxi^E$ and $\bxi^B$ are vectors from the vacuum to the electric and
magnetic singularities, respectively.  The dashed line is the 
CMS.\label{fig4}}

In finding the solutions to the BPS equations (\ref{EB}) we have
the discrete choice of charges at the sources, \ie\ the set
of integer charges $(Q^i_E,Q^i_B)$, which are constrained by
charge conservation (\ref{chrcons}).  In our simple example,
though, since only electric charges can flow into the
$\bX^E=(L,0)$ singularity and magnetic into the $\bX^B=(0,L)$ one,
we see that there are only two possible values for the
$\bxi^i$:
\be
\bxi^E \equiv \bX^E-\bX^0
\ee
for sources with only electric charges ($Q_B^i=0$), or
\be
\bxi^B \equiv \bX^B-\bX^0 
\ee
for sources with only magnetic charges ($Q_E^i=0$);
see figure~\ref{fig4}.  

This leads to a simplification of the BPS mass formula (\ref{M}) to
\be\label{M1}
M(Q_E,Q_B) = \sqrt{(Q_E)^2 \bxi^E\cdot\bxi^E + (Q_B)^2 \bxi^B\cdot\bxi^B
+ 2Q_E Q_B \bxi^E \times \bxi^B} .
\ee
In the case of purely electric or purely magnetic total charges, the
bound further simplifies to
\bea\label{M2}
M(Q_E,0) &=& |Q_E| |\bxi^E| ,\nonumber\\
M(0,Q_B) &=& |Q_B| |\bxi^B| ,
\eea
\ie\ the charge times the distance to relevant singularity on 
the Coulomb branch.

The CMS for the decay of a $(Q_E,Q_B)$ state to a $(Q_E,0)$ plus
a $(0,Q_B)$ state is then given by the curve on the Coulomb branch
for which $M(Q_E,Q_B) = M(Q_E,0) + M(0,Q_B)$.   {}From (\ref{M1}) and
(\ref{M2}) this implies that $|\bxi^E| |\bxi^B| = |\bxi^E \times \bxi^B|$, 
or that $\bxi^E$ be perpendicular to $\bxi^B$.  This describes a circle
on the Coulomb branch, shown as the dashed curve in figure~\ref{fig4}.

\subsection{Spike and prong solutions}

We will now illustrate some simple solutions to the BPS equations
and boundary conditions in our toy model.
To simplify the algebra and make the basic points clear we choose the
vacuum to be symmetrically placed at the point $\bX^0 = (X^0,X^0)$ on the
Coulomb branch. 

The simplest case is where there is a single charge source of
purely electric or magnetic charge.  For example, suppose $Q_B=0$.
By (\ref{EB}) and since
$\vB=0$ (because all the magnetic charges in the problem vanish) and
$\tan\alpha=-\xi_2^E/\xi_1^E$, the component of $\bX$ perpendicular to
$\bxi^E$ on the Coulomb branch must be constant, while its
component parallel to $\bxi^E$ satisfies Laplace's equation
with a source of total charge $Q_E$.  The solution is thus
simply
\be\label{Espike}
X_1 - X^0 = {Q_E\cos\beta \over4\pi|\vx-\vx_E|},\qquad
X_2 - X^0 = {Q_E\sin\beta \over4\pi|\vx-\vx_E|},
\ee
which implies electric and magnetic fields
\be
\vE = {Q_E \over 4\pi} {\vx -\vx_0 \over |\vx - \vx_0|^3},
\qquad \vB=0 .
\ee
Here $\beta$ is the angle on the Coulomb branch shown in
figure~\ref{fig4}, and $\vx_E$ is the spatial location of the electric
charge source.  Since both $\alpha$ and $\alpha+\pi$ satisfy
(\ref{tanal}) the sign of $Q_E$ in (\ref{Espike}) is undetermined; we
will determine it below. 

Following the brane picture of \cite{cm9708} we will call this a
``spike'' solution.  In the brane picture, the space-time coordinates
are identified with the worldvolume coordinates of a D3-brane, and
the $\bX$ scalars take values in the dimensions transverse to
the brane.  Thus the solution (\ref{Espike}) can be thought of as
describing a semi-infinite spike-like deformation of the brane in this
enlarged space.  As described in \cite{cm9708} such a solution can be
identified with a fundamental string ending on the D3-brane.
Likewise, the solution for a purely magnetically charged BPS state
is given by
\be\label{Bspike}
X_1 - X^0 = {Q_B\sin\beta \over4\pi|\vx-\vx_B|},\qquad
X_2 - X^0 = {Q_B\cos\beta \over4\pi|\vx-\vx_B|},
\ee
which can be interpreted as a spike with mass per unit length equal to
that of a D-string attached to the D3-brane.

We still need to apply our fuzzy ball boundary conditions
(\ref{dirbcE}) and (\ref{dirbcB}) to these solutions.  These boundary
conditions fix the undetermined signs of $Q_E$ and $Q_B$ in
(\ref{Espike}) and (\ref{Bspike}).  The signs of these solutions are
appropriate for
\be
X^0 \le L/2
\ee
or, equivalently, for $\beta > -\pi/4$, which we will assume from now
on.  The boundary conditions also determine the radii of the cutoff
spheres $S^2_\Lambda$ to be
\be
r_\Lambda = {|Q_E|\over 4\pi\ell}
\ee
implying a cutoff energy scale of $\Lambda \sim \ell$, where $\ell =
|\bxi^E| = |\bxi^B|$ is the distance from the vacuum to either Coulomb
branch singularity.  This is indeed the appropriate cutoff scale for
the low energy effective theory since there are new light charged
particles with masses (\ref{M2}) proportional to
this Coulomb branch distance.

The simple spike solutions found above do not have any interesting
structure on the Coulomb branch: they exist for any vacuum and obey
exactly Dirichlet boundary conditions with spherical boundaries which
we expected to be only approximately satisfied in general.  The
situation becomes more interesting when we turn to states with two or
more charge sources.

First consider a purely electrically charged two center solution with
$Q_E = Q_E^1 + Q_E^2$.  It is easy to find solutions with
the fuzzy ball boundary conditions (\ref{dirbcE}) and (\ref{dirbcB}).
Indeed,
\bea
X_1 - X^0 &=& {Q_E^1\cos\beta \over4\pi|\vx-\vx_{E,1}|}
+ {Q_E^2\cos\beta \over4\pi|\vx-\vx_{E,2}|},\nonumber\\
X_2 - X^0 &=& {Q_E^1\sin\beta \over4\pi|\vx-\vx_{E,1}|}
+ {Q_E^2\sin\beta \over4\pi|\vx-\vx_{E,2}|},
\eea
does the job as long as the distance
\be
r_{12} = |\vx_{E,1}-\vx_{E,2}|
\ee
between the two sources satisfies
\be\label{ineq}
r_{12} > {1\over 4\pi \ell} .
\ee
That is to say, as long as sources are sufficiently far apart one can
enclose the sources in disjoint spheres within which $\bX$ take the
value $(L,0)$.  This behavior is sensible: since a static
configuration of two charged BPS states with commensurate charge
vectors is itself BPS, there should exist a static low energy
configuration for any separation of the sources.  The restriction
(\ref{ineq}) that they not be too close just reflects the fact that
the low energy description of their interaction breaks down on scales
$r\lesssim \ell^{-1}$.

Now focus on a dyonic state with both electric and magnetic charges.  For
simplicity we take $(Q_E,Q_B)=(1,1)$.  By (\ref{tanal}) we see
that $\alpha=\pi/2$, so the BPS equations (\ref{EB}) reduce to
\be
\vE = \vn X_1 , \qquad \vB = \vn X_2.
\ee
{}From the linearity of the BPS
equations and since $\bX$ are harmonic functions, we expect the 
solution to be given at
least approximately by a solution of the form
\bea
X_1 - X^0 &=& {\cos\gamma\over4\pi|\vx-\vx_E|} 
+ {\sin\delta\over4\pi|\vx-\vx_B|} ,\nonumber\\
X_2 - X^0 &=& {\sin\gamma\over4\pi|\vx-\vx_E|}
+ {\cos\delta\over4\pi|\vx-\vx_B|} .
\eea
Compatibility with Gauss' law and the BPS equations then imply
that $\gamma=\delta=0$ so
\bea\label{soln}
X_1 - X^0 &=& {1\over4\pi|\vx-\vx_E|} ,\nonumber\\
X_2 - X^0 &=& {1\over4\pi|\vx-\vx_B|} .
\eea
We want to check whether there are values of the electric and
magnetic charge source centers, $\vx_E$
and $\vx_B$, for which this solution satisfies our fuzzy ball
boundary conditions (\ref{dirbcE}) and (\ref{dirbcB}).  What this asks
is that the $\bX$ values taken by this solution enter (and go
through) a small ball around $\bX^E=(L,0)$ as $\vx\to\vx_E$ and a
small ball around $\bX^B=(0,L)$ as $\vx\to\vx_B$.  As $\vx\to\vx_E$
the above solution approaches
\bea
X_1-X^0 &=& {1\over4\pi r_\Lambda} 
,\nonumber\\
X_2-X^0 &\simeq& {1\over4\pi r_{EB}} ,
\eea
where
\be
r_\Lambda = |\vx-\vx_E|,\qquad r_{EB} = |\vx_E-\vx_B|.
\ee
The $X_1 = L$ fuzzy boundary condition is achieved if the
spatial cutoff length scale around the electric source
is taken to be
\be
r_\Lambda = {1\over4\pi (L-X^0)},
\ee
while the $X_2=0$ boundary condition fixes the separation
of the electric and magnetic charge sources to be \cite{kk0002}
\be\label{sep}
r_{EB} = {1\over4\pi (-X^0)} .
\ee
Furthermore, on the sphere of radius $r_\Lambda$ around 
the electric source at $\vx_E$, by (\ref{soln}) $X_2$ takes values in 
a range $|X_2|<\widehat\Lambda$ with
\be\label{Lahat}
\widehat\Lambda = {(X^0)^2 L \over 1-2(X^0/L)}.
\ee
The boundary conditions at the magnetic source give the same
results.  

\EPSFIGURE{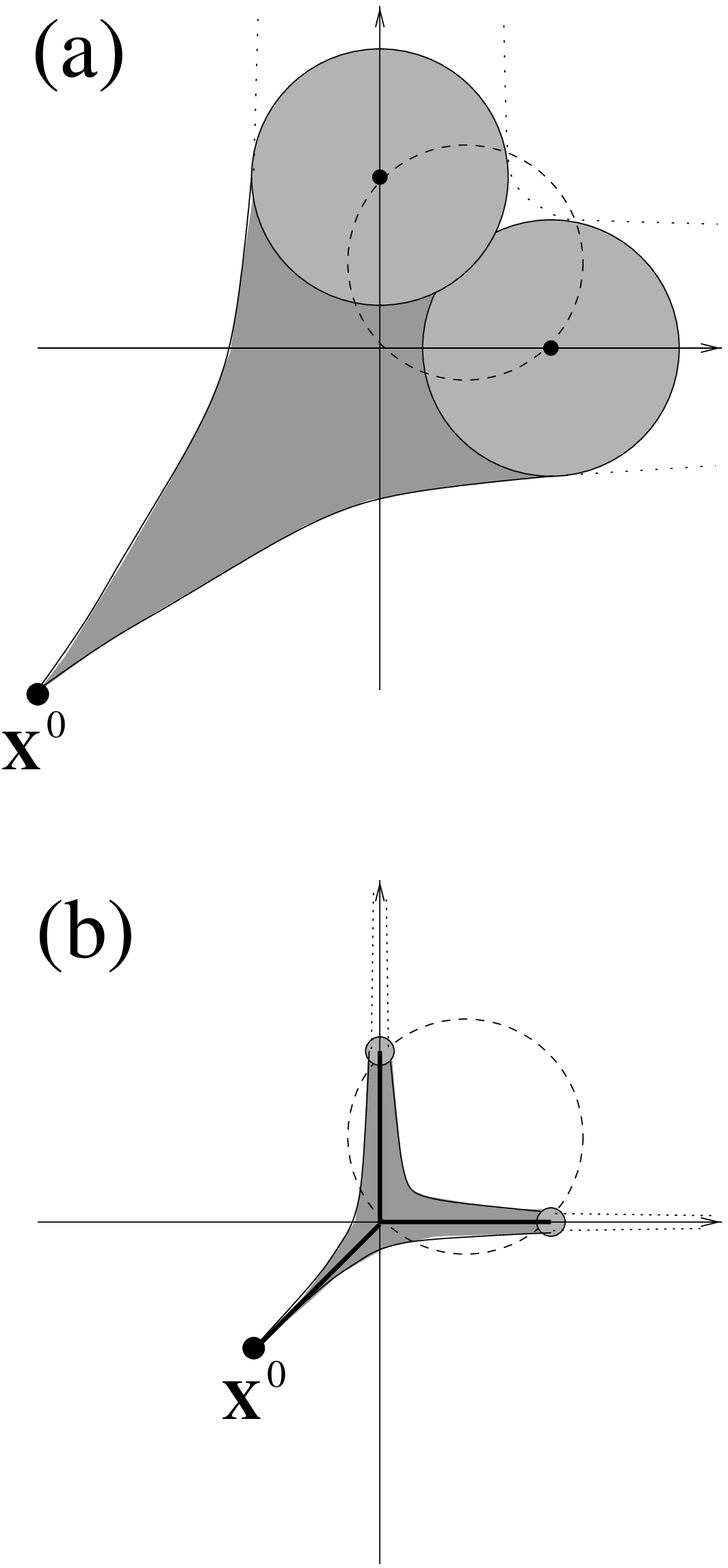,width=14em}{The shaded regions are the
images of the low energy dyon solution in the Coulomb branch for various
values of the vacuum $\bX^0$.  They are only valid outside the
lighter circles about the singularities which denote the $\widehat\Lambda$
fuzzy ball boundary conditions.\label{fig5}}

The conditions for our solution to be consistent are thus that
\be\label{bndry1}
X^0<0
\ee
since the distance $r_{EB}$ is necessarily positive, and that
$\widehat\Lambda$ is less than the distance $\sim L$ between the electric
and magnetic singularities on the Coulomb branch, which implies
\be\label{bndry2}
|X^0| \lesssim L .
\ee

\noindent
These conditions illustrate the qualitative picture of decaying
BPS states that we have argued for in the previous section.
Referring to figure~\ref{fig4} we see that $X^0=0$ is
the location of the CMS.  So (\ref{bndry1}) implies 
that a solution exists on one side of the CMS, and ceases to
exist once one crosses it.  Furthermore, the region of self-consistency
of the fuzzy ball boundary conditions includes the CMS;  indeed, the
fuzzy ball boundary conditions become more and more accurately simple
Dirichlet boundary conditions as one approaches the CMS since
(\ref{Lahat}) implies that
\be
\widehat\Lambda \sim (X^0)^2/L
\ee
vanishes as $X^0\to0$.  Finally, (\ref{sep}) shows that the spatial
distance between the charge cores diverges as we approach the CMS.
Figure~\ref{fig5} illustrates the qualitative behavior of these
solutions for different values of $X^0$.  These figures show
the projection of our low energy solutions on the Coulomb branch.
Note that this projection suppresses the spatial extent of the
field configuration; in particular note that as the vacuum approaches
the CMS, though the projection on the Coulomb branch degenerates to
a string web configuration, in space-time it expands as the positions
of the charge centers diverge.  Figure~\ref{fig5}(a)
shows how the fuzzy ball boundary conditions break down too far from
the CMS, while figure~\ref{fig5}(b) illustrates how the image of
the solution in the Coulomb branch more closely approximates
a string web configuration (shown as heavy lines) in the limit
as the vacuum approaches the CMS.  

We will refer to these multi-center solutions as brane prongs, again in
analogy to the brane picture where the Coulomb branch is realized as
additional transverse spatial dimensions.  The breakdown of the low
energy solution when $X^0$ is large compared to $L$ so that
(\ref{bndry2}) is not satisfied simply reflects the fact that in this
case the uncertainties in the locations of $\vx_E$ and $\vx_B$ are of
the order of the separation $r_{EB}$ itself, so now one cannot really
tell if there are two centers of charge in the low energy theory or
one.  Thus this configuration effectively looks like one brane spike
from a $(1,1)$ dyon.

Finally, it is straightforward to generalize the calculations of this
section to the case of arbitrary position of the $\bX^0$ of the
vacuum, arbitrary $(Q_E,Q_B)$ charge sector, and arbitrary value for
the low energy $\U(1)$ coupling $g$.  The result is a static solution
for the $(Q_E,Q_B)$ dyon outside the CMS for any relatively prime
$Q_E$ and $Q_B$, which decays into $Q_E$ charge $(1,0)$ plus $Q_B$
charge $(0,1)$ particles inside the CMS.  A two particle state with
mutually local charges (\ie\ commensurate charge vectors) is BPS and
so a static solution exists for any spatial separation.  Thus our low
energy method is not able to determine whether there is a one particle
bound state at threshold for electric and magnetic charges which are
not mutually prime.

\subsection{Recovery of a string junction picture}

We have thus explicitly verified all the qualitative features expected
of a BPS state decaying across a CMS.  The last thing to show is how
this information extracted from the low energy effective action on the
Coulomb branch is equivalent to a string web picture of the BPS
states.

The string web picture is recovered in two stages.  First, in the limit
as the vacuum approaches the CMS, the image of the low energy field
configuration in the Coulomb branch degenerates to a collection of
curves on that space.  These curves are the images of brane spike
solutions which carry precisely the energy per unit length of the
corresponding $(p,q)$ string \cite{cm9708}.  This is easy to see in our
example by taking $X^0\to0$ in (\ref{soln}) and comparing to our
electric and magnetic spike solutions (\ref{Espike}) and (\ref{Bspike})
(with $\beta=0$).  Thus arbitrarily close to the CMS the projection onto
the Coulomb branch of all the $(Q_E,Q_B)$ charged states look like
strings of appropriate tension stretched across the Coulomb branch. 

In string theory, 3-pronged string states that stretch between D-branes
satisfy charge conservation and tension balance. For example, the above
configuration corresponds to a $(1,0)$ (fundamental string), a $(0,1)$
(D-string)and a $(1,1)$ string stretched between three D3-branes.  The
three prongs or string legs meet at a common point, the junction, which
remains point-like even arbitrarily close to the CMS. In the field
theory picture discussed in this section, we have seen that the leg
corresponding to the decaying $(1,1)$ dyon grows in spatial size (\ie,
along the D3-brane worldvolume) near the CMS. Arbitrarily close to the
CMS, the configuration resembles two separate strings that end on the
D3-brane corresponding to the decaying $U(1)$. Thus there does not
appear to be a point-like junction as the decaying configuration
approaches the CMS in the field theory picture. 

The difference between the two perspectives is the result of different
orders of limits.  Consider the Dirac-Born-Infeld action for the
D3-brane corresponding to the decaying dyon, treating it as a probe in a
background of other D3-branes. The brane prong corresponding to the
decaying $(1,1)$ dyon then ends on other D3-branes that are treated as a
fixed background.  Looking at the quadratic terms and comparing with the
low energy effective action we have used above, we can see that the
scalars $X$ in the field theory with mass dimension unity and the
coordinates in the brane transverse space, $x$, are related by $X =
x/\alpha'$, where $\alpha'$ is the string length squared. The separation
between the charge centers in the D-brane worldvolume is
\be
r_{EB} \sim 1/(-X_0) =  \alpha' / (-x_0).
\ee
The low energy field theory is a good approximation in the $\alpha' \to
0$ limit holding the scalar vevs, including $X_0$, fixed.  On the other
hand, in the string junction/geodesic picture in string theory, the
coordinates $x_0$ are what are held fixed: taking $\alpha' \to 0$ to
suppress higher stringy corrections gives a vanishing separation
$r_{EB}$ between the charge centers, \ie\ a pointlike junction.  For
$x_0 = 0$, the separation is indeterminate, which corresponds to the two
strings ending anywhere on the D-brane.  This recovers the string theory
result and it further illustrates that the point-like junction is not
visible within the field theory approximation. 

The second stage is to extend this picture to the rest of the Coulomb
branch by continuity and matching to the BPS mass formula.  This can be
done uniquely since the given tension of a $(Q_E,Q_B)$ string (namely
$T(Q_E,Q_B) = \sqrt{Q_E^2+Q_B^2}$ in our example where we have set the
coupling $g=1$) fixes the path away from the CMS it must be extended in
in order to give the contribution to the state's mass required to match
its BPS mass.  That this path is a geodesic on the Coulomb branch
follows from the form of the BPS mass formula \cite{s9608}.  In the case
of our toy example, the Coulomb branch is flat so the geodesics are
straight lines and it is clear that the string web picture of the BPS
states extends without obstruction over the whole Coulomb branch. 

The result is therefore the prediction in our toy model that outside
the CMS the spectrum consists of all $(Q_E,Q_B)$-charged dyons with
$Q_E$ and $Q_B$ relatively prime, while inside the CMS only the $(\pm
1,0)$ and $(0,\pm1)$ states are in the spectrum.  Of course, this is
only a toy example.  We will perform essentially the same analysis in
the next section for the \Nf\ SYM theories.

\section{BPS states in \Nf\ $\SU(N)$ theories}\label{s4}

In this section we turn to an analysis of the static BPS field
configurations of given total electric and magnetic charges in the low
energy $\U(1)^{N-1}$ theory on the moduli space of an \Nf\ $\SU(N)$ SYM
theory.  The \Nf\ superalgebra allows both 1/2 and 1/4 BPS particle
states.  From the expression for the BPS mass, there are no CMS for
1/2 BPS states, but there are CMS for 1/4 BPS states to decay.

In section~\ref{s4p1} below we will effectively rederive the \Nf\ BPS
mass formula, as well as the BPS equations in the low energy theory.
We then find solutions to the BPS equations subject to our fuzzy ball
boundary conditions and show how they reproduce a string web picture
on the moduli space of the \Nf\ theory.

However this is not the usual string web picture of BPS states in \Nf\
theories derived from string theory.  In this picture \cite{w9510} the
$\U(N)$ SYM theory is realized by open strings and D1-branes ending on
a set of $N$ parallel D3-branes in a flat 10-dimensional IIB string
theory background.  The relative positions of the $N$ D3-branes in the
transverse 6-dimensional space form the $6N$-dimensional moduli space
of the \Nf\ theory.  BPS states are described by webs of $(p,q)$
strings stretching between the D3-branes \cite{b9712}; a web has
charge $(Q_E,Q_B)$ with respect to a given low energy $\U(1)$ factor
if it ends on the associated D3-brane with a $(Q_E,Q_B)$ string.
These string webs thus live in a 6-dimensional space stretched between
$N$ point sources, which is different from the picture we derive below
as string webs stretched in the $6N$-dimensional moduli space and
ending on $6(N-1)$-dimensional singular submanifolds.  In
section~\ref{s4p2} below we show how the 6-dimensional string theory
webs are obtained from our $6N$-dimensional webs by a simple mapping.
The basic reason this works is that the $6N$-dimensional moduli space
${\cal M}$ of the $\U(N)$ theory is ${\cal M} = (\R^6)^N/S_N$ where
the permutation group $S_N$ interchanging the $\R^6$ factors is the
Weyl group of $\U(N)$.  Because of this simple relation between the
transverse $\R^6$ of the D3-branes and $\cal M$, webs on $\cal M$ can
be uniquely mapped onto webs in $\R^6$, recovering the string picture.

A similar construction presumably also works for the $\SO$ and $\Sp$
\Nf\ SYM theories, since their Weyl groups are also fairly simple,
differing from the action of the $\SU$ Weyl group by the addition of
$\Z_2$ identifications on each $\R^6$.  This will fold webs on $\cal
M$ down to webs on $\R^6/\Z_2$, thus realizing the string web picture
of BPS states for these theories which are found by placing D3-branes
in the background of an appropriate orientifold O3-plane.
The action of the Weyl groups of the exceptional groups is more
complicated, and it is not clear whether any folding of our webs on
$\cal M$ to a 6-dimensional space can be performed.  This may
``explain'' why there are no D3 brane constructions of the \Nf\ SYM
theories with exceptional gauge groups.

\subsection{BPS states as webs on the moduli space}\label{s4p1}

The generic point on the moduli space of an \Nf\ SYM theory with gauge
group $\SU(N)$ is described by a low energy $\U(1)^{N-1}$ \Nf\ theory.
A convenient trick to simplify the algebra will be look at the
$\U(N)$ theory instead, which has a $\U(1)^N$ effective description,
but restrict ourselves to states which are neutral under the extra
$\U(1)$.

Now, each $\U(1)$ \Nf\ multiplet has six real scalar fields, so the
moduli space is locally coordinatized by the values of the $6N$ scalars
$X_{r a}$ where $r=1,\ldots,N$ and $a=1,\ldots,6$.  The other bosonic
massless fields are the electric and magnetic fields $\vE_r$ and $\vB_r$
for each $\U(1)$ factor.  We are interested in static low energy
solutions carrying electric and magnetic charges $(Q_{Er}, Q_{Br})$ with
respect to these fields.  We normalize the fields so that the low energy
effective action for the bosonic fields is
\be
S= \int d^4x\, \sum_{r=1}^N \left( {1\over4g^2} F_{r\,\mu\nu}
F^{\mu\nu}_r + {\vartheta\over64\pi^2} \epsilon_{\mu\nu\rho\sigma}
F_r^{\mu\nu}F_r^{\rho\sigma} + {1\over4g^2} \sum_{a=1}^6 
\del_\mu X_{r a} \del^\mu X_{r a} \right),
\ee
with the $\vE_r$ and $\vB_r$ related to $F^{\mu\nu}_r$ in the usual way.
The fact that all the $\U(1)$'s have the same coupling and no
cross-couplings reflects the specific basis of $\U(1)^N$ we have chosen.
The fact that the coupling is independent of the $X_{r a}$ reflects the
flatness of the metric on $\cal M$ for \Nf\ theories.  We normalize the
charges so that
\bea
\oint_{S^2} \vE_r\cdot d\vec a &=& g^2 Q_{Er},\nonumber\\
\oint_{S^2} \vB_r\cdot d\vec a &=& g^2 Q_{Br}.
\eea
Thus, the Dirac quantization condition plus the effect of the
theta angle imply that the charges are quantized as
\be
Q_{Er} = n_{Er} + {\vartheta\over 2\pi} n_{Br}
\qquad \mbox{and} \qquad
Q_{Br} = {4\pi\over g^2} n_{Br}
\ee
for some integers $n_{Er}$ and $n_{Br}$.

\subsubsection{Moduli space and boundary conditions}

Globally, the moduli space is
\be
{\cal M} = \R^{6N}/S_N
\ee
where $S_N$ acts by permuting the $N$ $\R^6$ factors.  Suppose we have
some ordering of points in $\R^6$ (say ``alphabetical'' ordering on
their six coordinates in some given basis) denoted by $X_{1a} \ge
X_{2a}$ where $X_1$ and $X_2$ are two points in $\R^6$.  Then we can
take as a fundamental domain of $S_N$ on $\R^{6N}$ the set of all $N$
6-vectors satisfying
\be\label{wedge}
X_{1a} \ge X_{2a} \ge \cdots \ge X_{Na} .
\ee
The boundaries of this wedge-like convex domain are identified by 
the $S_N$ action.

The locus of fixed points of the action of $S_N$ are orbifold
singularities of $\cal M$ where BPS states of certain charges become
massless.  The typical such singularity is the fixed line under
interchange of, say, $X_{1a}$ with $X_{2a}$.  The result is a
$6(N-1)$-dimensional manifold of $\Z_2$ singularities at
$X_{1a}=X_{2a}$, with all the other $X_{r a}$ arbitrary for
$r>2$.  At this singularity states with charges $Q_{E1}=-Q_{E2}$
and $Q_{B1}=-Q_{B2}$ arbitrary and $Q_{Er}=Q_{Br}=0$ for
$r>2$ become massless.  Actually all such $6(N-1)$-dimensional
manifolds of $\Z_2$ singularities are identified by the $S_N$
orbifolding.  However, which charged states become massless
there depends on the direction the singularity is approached from.
Thus it is convenient to treat $\cal M$ as the wedge (\ref{wedge})
with different charge images of the $\Z_2$ singularity on 
its boundary.

In addition, these images of the $\Z_2$ singularities may intersect
one another along submanifolds of smaller dimension where states
charged under three or more $\U(1)$ factors become massless.  In fact,
on the 6-dimensional submanifold of $\cal M$ where all the $X_{r
a}$, $r = 1,\ldots,N$ are equal, BPS states of arbitrary charge
become massless.  Since we are only considering states which are
neutral with respect to the diagonal $\U(1)$ factor (to decouple the
extra $\U(1)$ of the $\U(N)$ gauge group), all configurations will be
invariant under translations in $\cal M$ by any such vector of equal
$X_r$'s.

With this description of the moduli space and its singularities in
hand, we are ready to solve for static low energy field configurations
in a given charge sector subject to our fuzzy ball boundary
conditions.  Since we are looking for BPS configurations which
preserve at least 1/4 of the supersymmetries, we can 
restrict ourselves to configurations which lie in some fixed
two-dimensional subspace of each $\R^6$ factor, which we will
take to be that given by the $X_{r a}$ with $a=1,2$.  Furthermore, 
the residual (${\cal N}=1$) supersymmetry implies that the BPS
configurations must depend holomorphically on the $X_{r 1} 
+ i X_{r 2}$ complex coordinates.
It will thus be convenient to introduce a notation in which
all quantities are assembled into $N$ component complex vectors
defined by
\bea
X_r &\equiv& X_{r 1}+iX_{r 2}, \nonumber\\
\vF_r &\equiv& \vE_r + i\vB_r,\nonumber\\
Q_r &\equiv& Q_{Er}+iQ_{Br}.
\eea

Denote the total charge of our configuration by $Q^0_r$.  Suppose our
configuration has $M$ charge centers with (approximate) spatial
positions $\vx=\vx_i$, with charges labeled by $Q^i_r$,
$i=1,\ldots,M$.  (We will mainly concentrate on the case of prong
solutions with $M=2$ below.)  Then charge conservation implies
\be\label{chrgcon}
\sum_{I=0}^M Q_r^I = 0.
\ee
Label the boundaries of space by the spheres $S^2_I$ where $S^2_0$
refers to the sphere at infinity, while the $S^2_i$ for $i=1,\ldots,M$
are small spheres around the $M$ charge sources at $\vx=\vx_i$.  The
Gauss' law implies
\be\label{nfgauss}
\oint_{S^2_I} d{\vec a} \cdot {\vF_r} = g^2 Q^I_r .
\ee
The scalars approach
\be\label{nfinfbc}
X_r = X^0_r\quad\mbox{on $S^2_0$},
\ee
where $X^0_r$ are the moduli space coordinates of the vacuum.  Our
fuzzy ball boundary conditions imply that the $X_r$ will approach (at
least approximately) constant values on the $S^2_i$ boundaries, which
we will denote
\be\label{nfdirbc}
X_r \simeq X_r^i \quad\mbox{on $S^2_i$}.
\ee
Furthermore, the $X^i_r$ are constrained to lie in only those singular
submanifolds of $\cal M$ where states of charge $Q_r^i$ become
massless.  {}From the above description of $\cal M$ the $X_r^i$ must
live on the appropriate singular submanifold where a BPS state
with its associated charges becomes massless.

\subsubsection{BPS bound}

The mass of our configuration is computed by integrating
the field energy density (with an implicit sum over repeated
indices)
\be
M ={1\over2g^2} \int d^3\vx\, (\vF^*_r\cdot\vF_r 
+ \vn X^*_r\cdot \vn X_r ).
\ee
This can be rewritten as
\be \label{decomp}
M= {1\over2g^2} \int d^3\vx\, |\vF_r - A_{rs} \vn X_s|^2 + 
{1\over 2g^2}\Rea \int d^3\vx\, A_{rs} \vF^*_r\cdot \vn X_s 
\ee
if $A_{rs}$ is an $U(N)$ matrix
\be\label{A}
A^*_{rs} A_{rt} = \delta_{st} .
\ee
Using $\vn\cdot \vF_r=0$ away from the sources, the
second term in (\ref{decomp}) becomes a surface term which
can be evaluated using (\ref{nfgauss})--(\ref{nfdirbc}) to 
\be\label{massexp}
M = {1\over2g^2} \int d^3\vx\, |\vF_r - A_{rs} \vn X_s|^2 + 
\Rea \sum_I Q_r^{I*} A_{rs} X_s^I .
\ee
Since the first term on the right hand side
is positive definite, we get the bound 
\be\label{bnd}
M \geq \Rea \{Q_r^{I*} A_{rs} X_s^I\}.
\ee
The BPS bound arises from maximizing the right hand side of
this expression subject to (\ref{A}) 
\be\label{nfbps1}
M_{BPS} = \max_{A\in\U(N)} \Rea \{Q_r^{I*} A_{rs} X_s^I \} .
\ee
Finally, this BPS bound is saturated when the BPS equations
\be\label{nfbpseq}
\vF_r = A_{rs} \vn X_s, \qquad \nabla^2 X_s = 0,
\ee
are satisfied for the $A_{rs}$ which maximizes (\ref{nfbps1}), 
where the last equation is a result of the divergence free
property of $\vF_r$.

The expression for $M_{BPS}$ can be simplified slightly using charge 
conservation.  Define by
\be
\xi_s^i \equiv X_s^i - X_s^0 
\ee
the vector pointing from the vacuum $X_s^0$ to the source (singularity)
at $X_s^i$.  Then, by (\ref{chrgcon}),
\be
Q_r^{I*} A_{rs} X_s^I = Q_r^{i*} A_{rs} \xi_s^i ,
\ee
and the BPS bound becomes
\be\label{nfbps2}
M_{BPS} = \max_{A\in\U(N)} \Rea \{Q_r^{i*} A_{rs} \xi_s^i \}.
\ee

Now, as described above, the singularities of $\cal M$ are whole
submanifolds, not isolated points, so the vectors $\xi_s^i$ pointing
to the singularities can vary continuously.  We must therefore vary
our boundary conditions to find the lightest (potentially) BPS state
in the given charge sector.  This has two consequences.  First, this
implies that generically the charge sources will lie along the the
$\Z_2$ singularities in $\cal M$, \ie\ that only with some special
fine tuning will a source charged under three or more $\U(1)$'s not
split into several sources each with (opposite) charges under only two
$\U(1)$'s.  This follows simply because the multi-charged sources are
constrained to lie in submanifolds at the intersection of the
higher-dimensional $\Z_2$ singularities.  The second consequence is
that the true BPS bound is given by the the minimization of
(\ref{nfbps2}) with respect to the set of vectors $\xi^i_r$ ending on
the $\Z_2$ singularities
\be\label{nfbps3}
M_{BPS} = \min_{\xi\in{\rm sings.}} \max_{A\in\U(N)} 
\Rea\{Q_r^{i*} A_{rs} \xi_s^i \}.
\ee

It is important to note that the maximization with respect to the
rotation $A_{rs}$ be done for each $\xi^i_r$ {\em before} the
minimizing with respect to $\xi_r^i$.  If the space of singularities
over which the $\xi^i$ can vary were compact and smooth, then (since
the space of $A$'s is compact) there would be no issue over the order
in which the extremizations were performed.  But the space of
singularities over which the $\xi^i$ can vary themselves have
singularities, and so some care must be taken in case the solution
lies not at a saddle point of $\Rea Q^{*i}\cdot A\cdot \xi^i$, but at
one of these singularities.  We have found, however, that for generic
$Q^i$'s and $X^0$ this does not happen so the order of extremization
can be reversed.  This greatly simplifies the problem of finding
$M_{BPS}$ in \Nf\ theories because the singularities lie along linear
subspaces of the flat moduli space.  Thus minimization with respect to
the $\xi^i$ by itself gives $M$ independent complex conditions on $A$,
essentially fixing it and $M_{BPS}$ entirely in the cases of one or
two charge sources ($M=1$ or 2) each charged under only two $\U(1)$
factors.  This minimization is carried out in the Appendix for two 
charge sources.

Even in this case one must still do the maximization with respect the
$A$ to determine the values of the $\xi^i$ in terms of given source
charges $Q_s^i$ and vacuum $X_s^0$.  Solving this minimization and
maximization problem for the $\xi^i$ and $M_{BPS}$ with an arbitrary
number of given charge sources $Q_r^i$ is difficult in general.
However, as we will explain below, it is sufficient to solve it for just
two sources each charged under only two $\U(1)$ factors.  This is also
done explicitly in the Appendix for two charge sources.  We see
there that the $\xi^i$ are not completely determined by the
extremization of the BPS bound, but are ambiguous up to a single
undetermined real parameter.  The final condition needed to fix
the boundary conditions (the $\xi^i$) comes from demanding that
solutions to the BPS equations exist.

\subsubsection{BPS solutions}

So, suppose we fix the $\xi^i$, and thus the boundary values of the
scalars.  We will now solve the low energy BPS equations (\ref{nfbpseq})
subject to the charge boundary conditions (\ref{nfgauss}), the boundary
conditions at infinity (\ref{nfinfbc}), as well as our fuzzy ball
boundary conditions centered around the above determined boundary
values.  In doing so we will determine necessary conditions for a
solution to exist given these boundary conditions.  These conditions can
phrased as the conditions that a certain matrix $\alpha_{ij}$ be real,
symmetric, and have only positive entries.  The reality and symmetry
condition will provide the extra condition needed to determine the
$\xi^i$ for two or more charge sources.  Finally, the positivity
condition is satisfied only on one side of the CMS, and so it is this
condition which determines the stability of BPS states. 

The BPS equations can be solved as in the toy model of section~\ref{s3}
by a superposition of single source solutions:
\be\label{nfsoln3}
\xi_s(\vx) = \sum_{i=1}^M {Q_r^i A^*_{rs} \over 4\pi |\vx - \vx_i|} ,
\ee
where we have defined
\be
\xi_s(\vx) \equiv X_s(\vx) - X_s^0 .
\ee
The numerators on the right hand side are determined by (\ref{nfbpseq})
and (\ref{nfgauss}).  Now, the fuzzy ball boundary conditions are that
$\xi_s$ goes through a small ball around $\xi_s^i$ as $\vx\to\vx_i$.  In
this limit $\xi(\vx)$ approaches
\be
\lim_{\vx \to \vx_i} \xi_s \simeq {A^*_{rs} Q_r^i \over \epsilon} + 
\sum_{j \neq i}^M {A^*_{rs} Q_s^j \over r_{ij}} \to \infty
\ee
where 
\be\label{defepr}
\epsilon \equiv 4\pi |\vx - \vx_j|, 
\quad \mbox{and}\quad
r_{ij} \equiv 4\pi |\vx_i - \vx_j|.
\ee
Thus as $\vx\to\vx_j$ our solutions go to infinity asymptoting a line
along the $A^*_{rs} Q_s^j$ direction and with intercept
$\delta^i_r=\sum_{j \neq i} {A^*_{rs} Q_s^j \over r_{ij}}$.
So a necessary condition for the fuzzy ball boundary
conditions to be satisfied is that the approximate boundary value
$\xi^i$ at the $i$th source lies on this asymptote:
\be\label{nfbc1}
\xi^i_r = \alpha A^*_{rs} Q_s^i + 
\sum_{j \neq i} {A^*_{rs} Q_s^j\over r_{ij}},
\qquad \mbox{for some}\ \alpha > 0.
\ee

In particular, the fact that $r_{ij} >0$ means that the condition
(\ref{nfbc1}) has the form
\be\label{defalf}
\xi^i = \sum_j \alpha_{ij} A^*_{rs} Q_s^j
\ee
for $\alpha_{ij}$ a real symmetric matrix of positive numbers.
Multiplying this equation on both sides by the unitary matrix $A_{rt}$
implies therefore that a necessary condition for there to exist a
solution to the BPS equations with our fuzzy ball boundary conditions is
that {\em the $A_{rs}\xi_s^i$ must be in the real, symmetric, and
positive span of the $Q_r^k$:}
\be\label{defalf2}
A_{rs}\xi^i_s = \sum_k \alpha_{ij} Q^j_r ,
\qquad \alpha_{ij}=\alpha_{ji}\ge0.
\ee

\subsubsection{Spike solutions (1/2 BPS states)}

It is straightforward to find single source brane spike solutions
representing 1/2 BPS states in the low energy theory.  These
correspond to solutions with a single charge source charged under one
pair of $\U(1)$ factors (recall that the total charge under the diagonal
$\U(1)$ is assumed zero to decouple it).  Thus, without loss of
generality, we can restrict ourselves to just two $\U(1)$ factors, with
everything neutral under the diagonal $\U(1)$.  We will take the
associated complex charge vector to be
\be
Q = \pmatrix{q \cr -q \cr}.
\ee
We also choose the vacuum to be at the complex scalar vevs
\be\label{nfspike1}
X^0 = \pmatrix{a \cr -a \cr}.
\ee
{}From our earlier discussion, the singular submanifold of the (relevant
one-complex-dimensional submanifold of) moduli space where states of
charge $Q$ can end is only the origin
\be
X^{\rm sing} = \pmatrix{0\cr 0\cr}.
\ee
This follows because the singular manifolds are those points with equal
complex coordinates under the two $\U(1)$'s; but the decoupling of the
diagonal $\U(1)$ restricts us to the submanifold where the two
coordinates sum to zero, as in (\ref{nfspike1}). 

Since we are effectively restricted to a one-complex-dimensional space,
it is convenient to change basis on the moduli space by means of a
simple unitary transformation
\be
\pmatrix{a\cr -a\cr} \to
{1\over\sqrt2}\pmatrix{1 & -1\cr 1 & 1\cr}
\pmatrix{a\cr -a\cr} = \pmatrix{\sqrt2 a \cr 0\cr} .
\ee
In this basis, where we can ignore the last component which will always
vanish by the tracelessness condition, the charge and vacuum are simply
the complex numbers
\be
Q = \sqrt2 q,\qquad X^0 = \sqrt2 a.
\ee

Our general expression derived above for the BPS mass (\ref{nfbps3})
simplifies to
\be\label{nfspmax}
M_{BPS} = \max_{\phi} \Rea\{Q^* e^{i\phi} (-X^0) \}
= \max_{\phi} \Rea\{ -2q^* a e^{i\phi} \}
= 2|q|\,|a| .
\ee
Here we have written the $\U(1)$ ``matrix" $A$ as the phase $e^{i\phi}$.
The maximization determines $\phi$ to be the phase of $-qa^*$. 

The solution to the BPS equations (\ref{nfsoln3}) also simplifies to
\be\label{nfspikesoln}
X(\vx) -a = {qe^{-i\phi} \over 4\pi |\vx - \vx_0|}
\ee
where, $\vx_0$ is the arbitrary spatial location of the charge source.
This solution clearly exists everywhere on the moduli space (\ie\ for
all $a$): it has no structure and satisfies Dirichlet boundary
conditions $X=0$ on the sphere around $\vx_0$:
\be\label{nfsph}
|\vx-\vx_0| = -{qe^{-i\phi} \over 4\pi a} = {|q|\over4\pi|a|}, 
\ee
where the last equality follows from the solution for $\phi$ in
(\ref{nfspmax}).  These properties are just as in the spike solutions in
our toy example of section~\ref{s3}. 

\subsubsection{Prong solutions (1/4 BPS states)}\label{s4p1prong}

Now let us specialize to the case $M=2$, that is, two sources, each 
charged under one pair of $\U(1)$ factors (recall that the total charge
under the diagonal $\U(1)$ is assumed zero to decouple it).  Thus,
without loss of generality, we can restrict ourselves to just three
$\U(1)$ factors, with everything neutral under  the diagonal
$\U(1)$.  We will take the associated complex charge vectors
to be
\be\label{nfQdef}
Q^1 = \pmatrix{q_1 \cr -q_1 \cr 0\cr},
\qquad
Q^2 = \pmatrix{0 \cr q_2 \cr -q_2\cr}.
\ee
We also choose the vacuum to be at the complex scalar vevs
\be
X^0 = \pmatrix{a \cr b \cr c\cr}, \qquad
a+b+c=0.
\ee
{}From our earlier discussion, the singular submanifold of the moduli
space where states of charge $Q^1$ and $Q^2$ can end can have
coordinates
\be
X^1 = \pmatrix{x \cr x \cr -2x\cr},
\qquad
X^2 = \pmatrix{-2y \cr y \cr y\cr},
\ee
respectively, for arbitrary complex $x$ and $y$.

In the Appendix we analyze the conditions on $x$ and $y$ that result
from extremizing the BPS bound and demanding reality and symmetry of the
$\alpha_{ij}$.  These conditions fix $x$ and $y$ completely. 

The result can be expressed as follows.  First we set up a convenient
notation.  Define $\theta_1$ and $\theta_2$ to be the phases of $q_1$
and $q_2$:
\be
q_j = |q_j| e^{i\theta_j} .
\ee
Then decompose each of the complex numbers $a$, $b$, $c$, $x$, $y$ in
the (non-orthogonal) basis $\{\bee_1,\bee_2\}$ defined by
\be
\bee_j \equiv e^{i(\theta_j-\phi)},
\ee
where $\phi$ is a phase to be defined below.  Thus, for example, we
write
\be
a = a_1 \bee_1 + a_2 \bee_2
\ee
for unique real numbers $a_j$; define similarly the real numbers $b_j$,
$c_j$, $x_j$,and $y_j$, for $j=1,2$. 

Finally, $\phi$ is defined by
\be\label{nfBPS5}
q_1 (b^* -a^*) + q_2 (c^* -b^*) = M_{BPS} e^{i\phi}.
\ee
In other words, $\phi$ is the phase of the left hand side, and
$M_{BPS}$, the BPS mass, is the norm:
\be\label{nfBPS6}
M_{BPS} = |q^*_1 (a-b) + q^*_2 (b-c)| .
\ee
Then the result for $x$ and $y$ is given by
\bea
x &=& -{1\over2}c_1 \bee_1 + a_2 \bee_2,\nonumber\\
y &=& c_1 \bee_1 - {1\over2} a_2 \bee_2.
\eea

The final condition for the existence of a prong type solution in the
\Nf\ effective action with our boundary conditions is that
\be\label{nonneg}
\alpha_{ij} \ge 0,
\ee
where the $\alpha_{ij}$ are defined in (\ref{defalf}).  {}From the
expressions for the $\alpha_{ij}$ derived in the Appendix it is easy to
see that (\ref{nonneg}) is equivalent to the four conditions
\bea\label{poscond}
b_1 \ge a_1, &\qquad& b_1 \ge c_1,\nonumber\\
c_2 \ge b_2, &\qquad& a_2 \ge b_2.
\eea
Note that (since $a+b+c=0$) the only way {\em both} conditions in either
line can be saturated is if $a_j=b_j=c_j=0$, a singular vacuum.  Thus
the non-singular way these conditions can be saturated is if only one
inequality in each line of (\ref{poscond}) is saturated, \eg,
\bea\label{nfCMS1}
b_1 > a_1, &\qquad& b_1 = c_1,\nonumber\\
c_2 > b_2, &\qquad& a_2 = b_2.
\eea
A check of our picture is that this boundary at which prong solutions
cease to exist corresponds to being on a CMS. 

We can check that this is indeed the case as follows.  Now,
(\ref{nfBPS5}) and (\ref{nfBPS6}) imply that the BPS mass can be written
as
\be\label{nfBPS7}
M_{BPS} = \Rea \left\{ e^{i\phi} \left[
q^*_1(b-a) + q^*_2 (c-b) \right]\right\},
\ee
which when expanded in the $\bee_i$ basis gives
\be\label{nfBPS8}
M_{BPS} = |q_1|\left[ (b_1-a_1) + (b_2-a_2)\cos(\theta_1-\theta_2) \right]
+ |q_2|\left[ (c_2-b_2) + (c_1-b_1)\cos(\theta_1-\theta_2) \right].
\ee
If, on the other hand, there is just one charge source of charge
$Q^1$ then its BPS mass is given by (\ref{nfBPS6}) with $q_2=0$,
implying
\be\label{nfBPSh1}
M_1 = |q_1|\, |b-a|
= |q_1| \sqrt{(b_1-a_1)^2 + (b_2-a_2)^2 
+ 2 (b_1-a_1)(b_2-a_2)\cos(\theta_1-\theta_2)} .
\ee
Similarly, the BPS mass for the charge $Q^2$ one-source state
\be\label{nfBPSh2}
M_2 = |q_2| \sqrt{(c_1-b_1)^2 + (c_2-b_2)^2 
+ 2 (c_1-b_1)(c_2-b_2)\cos(\theta_1-\theta_2)} .
\ee
The condition to be on a CMS is, by definition,
\be
M_{BPS} = M_1 + M_2,
\ee
which by (\ref{nfBPS8}), (\ref{nfBPSh1}), and (\ref{nfBPSh2}),
precisely corresponds to (\ref{nfCMS1}) as required.

Note that if the phases of the charge vectors are identical, \ie\
$\theta_1 = \theta_2$, then the vector space spanned by the $e_i$ basis
degenerates and there are no $1/4$ BPS states, only $1/2$ BPS states, 
as expected.

It will be useful to note that in the vicinity of the CMS and on the
side where prong solutions exist, \ie\ where (\ref{poscond}) is 
satisfied, the following inequalities hold:
\be\label{order}
b_1 > c_1 > 0 > a_1, \qquad\mbox{and}\qquad
c_2 > 0 > a_2 > b_2.
\ee
These follow from $a+b+c=0$, (\ref{poscond}), and the condition that we
are near the CMS, \ie\ $b_1\simeq c_1$ and $a_2\simeq b_2$. 

Putting all the pieces of this calculation together we have thus shown
that in the \Nf\ low energy effective theory, two source ``prong''
solutions (\ref{nfsoln3}) always exist on one side of the relevant CMS.
Furthermore, the CMS condition (\ref{nfCMS1}) is equivalent to
$\alpha_{12}=\alpha_{21}=0$ in (\ref{defalf}), which, comparing to
(\ref{nfbc1}), implies $r_{12} = \infty$ where $r_{12}$ is the spatial
source separation (\ref{defepr}).  Thus we learn that in the prong
solutions the sources have definite spatial separation which diverges
as we approach the CMS, in accord with the qualitative picture described
in section~\ref{s2}.  

Finally, in the limit as we approach the CMS the prong solutions obey
Dirichlet boundary conditions (as opposed to the more general fuzzy ball
boundary conditions) to ever-greater accuracy, just as in the discussion
in the toy example in section~\ref{s3}.  Hence there will always be some
distance from the CMS on the moduli space within which our low energy
fuzzy ball boundary conditions are self-consistent. 

\subsubsection{Webs on moduli space}

The final step in deriving a string web picture of BPS states on the
moduli space of the \Nf\ theory is to project the solutions we have
constructed above onto that moduli space. 

We start with the spike (or single source) solutions given in
(\ref{nfspikesoln}).  Since the spatial dependence only enters in the
positive factor $|\vx-\vx_0|^{-1}$, the image of $X(\vx)$ on the moduli
space is simply a straight line segment, starting at $X^0=\sqrt2 a$ (at
$\vx=\infty$) and ending at the origin $X=0$ on the sphere of
(\ref{nfsph}).  This, together with its mass (\ref{nfspmax}), is thus
consistent with its interpretation as a string of tension $|Q|=|\sqrt2
q|$ stretched a length $|\sqrt2 a|$ on the moduli space. 

A similar picture applies to the brane prong (or two source) solutions
as well.  As the vacuum approaches the CMS, the image of
the solution (\ref{nfsoln3}) in the moduli space approaches that
of a web of line segments, just as discussed of section~\ref{s3}.
One segment leads from the vacuum to the CMS, where two more
segments emanate, leading to the singularities $X^j$.  {}From
the form of the solution (\ref{nfsoln3}), the two segments
leading towards the singularities point along the direction
\be
Q_r^j A^*_{rs}= e^{-i\phi} Q^j_s
\ee
in the moduli space.  Recalling our definitions (\ref{nfQdef}) of
the complex charge vectors $Q^j$, it follows that the directions
of the line segments going to $X^1$ and $X^2$ can be written as
\be
|q_1| \pmatrix{\bee_1\cr -\bee_1\cr 0\cr},\qquad\mbox{and}\qquad
|q_2| \pmatrix{0\cr \bee_2\cr -\bee_2\cr},
\ee
respectively.  Then from our solution for the $X^j$ it follows
immediately that the three line segments that the prong solution
degenerates to are given by
\bea\label{nfstringweb}
X^0 \to \mbox{CMS}: \qquad X(t) &=&
\pmatrix{a\cr b\cr c\cr} 
+ t \pmatrix{(b_1-c_1)\bee_1\cr 
(c_1-b_1)\bee_1+(a_2-b_2)\bee_2\cr 
(b_2-a_2)\bee_2\cr}, \nonumber\\
X^1 \to \mbox{CMS}: \qquad X(t) &=&
\pmatrix{-{1\over2}c_1\bee_1+a_2\bee_2\cr 
-{1\over2}c_1\bee_1+a_2\bee_2\cr 
+c_1\bee_1-2a_2\bee_2\cr} 
+ t \pmatrix{-{3\over2}c_1\bee_1\cr 
+{3\over2}c_1\bee_1\cr 0\cr}, \\
X^2 \to \mbox{CMS}: \qquad X(t) &=&
\pmatrix{-2c_1\bee_1+a_2\bee_2\cr 
+c_1\bee_1-{1\over2}a_2\bee_2\cr 
+c_1\bee_1-{1\over2}a_2\bee_2\cr} 
+ t \pmatrix{0\cr +{3\over2}a_2\bee_2\cr 
-{3\over2}a_2\bee_2\cr}, \nonumber
\eea
where in all cases $t\in[0,1]$.  Notice that the
common $t=1$ endpoint of each of these segments,
\be
\pmatrix{-2c_1\bee_1+a_2\bee_2\cr
+c_1\bee_1+a_2\bee_2\cr 
+c_1\bee_1-2a_2\bee_2\cr}, 
\ee
satisfies (\ref{nfCMS1}) so is indeed on the CMS.

Finally, we check that this three-string junction on the moduli space of
the \Nf\ theory indeed gives the correct BPS mass just by adding the
lengths of its segments weighted by the tension of each string (which is
the norm of its charge vector in our conventions).  Showing this shows
that this string-junction picture can be continued to all vacua, and not
just those close to the CMS. 

The charges for the $X^j \to \mbox{CMS}$ segments are $Q^j$, $j=1,2$
respectively, thus the tension of these strings are $|Q^j| = \sqrt2
|q_j|$.  The multiplied by the lengths of their respective segments
gives their masses as
\bea\label{segma1}
M(X^1\to\mbox{CMS}) &=& 3|q_1|c_1,\nonumber\\
M(X^2\to\mbox{CMS}) &=& -3|q_1|a_2,
\eea
where we have used the fact that $c_1$ is positive and
$a_2$ negative from (\ref{order}).  The charge of the
$X^0 \to \mbox{CMS}$ segment is the total charge
which can be written
\be
Q^1+Q^2 = e^{i\phi} \pmatrix{+|q_1|\bee_1\cr
-|q_1|\bee_1+|q_2|\bee_2\cr -|q_2|\bee_2\cr} .
\ee
Now, as shown in the Appendix the condition determining $\phi$ 
implies eq.~(\ref{phicond}) which can be written as
\be
(a_2-b_2) = \beta |q_2|, \qquad  (b_1-c_1) = \beta |q_1| ,
\qquad\mbox{for some real}\ \beta >0,
\ee 
where we have again used the fact that $a_2-b_2$ and $b_1-c_1$ are
positive from (\ref{order}).  This implies that $e^{-i\phi}$ times the
charge vector and the vector on moduli space for the $X^0\to\mbox{CMS}$
line segment are proportional with positive real proportionality factor
$\beta$.  Thus the product of the lengths of these two vectors is the
same as their inner product, giving the mass of this string segment as
\be\label{segma2}
M(X^0\to\mbox{CMS}) = 
(b_1-c_1) \left[2|q_1| - |q_2|\cos(\theta_1-\theta_2)\right] +
(a_2-b_2) \left[2|q_2| - |q_1|\cos(\theta_1-\theta_2)\right].
\ee
Adding the masses of the segments (\ref{segma1}), (\ref{segma2}),
indeed gives the BPS mass (\ref{nfBPS8}).

This completes our construction of our representation of BPS states as
three-string junctions on the moduli space of \Nf\ theories. 

So far we have only dealt with one and two source solutions.  The
general BPS state will be charged under more than just three $\U(1)$
factors, and so will generically break up into more that just two charge
sources.  However, our low energy method can deal with such cases only
in an indirect manner.  The reason is that our approximate boundary
conditions are only valid when the vacuum is close to a CMS.  But a
generic solution with three or more charge sources generically has two
or more distinct CMS corresponding in the string web picture to two or
more separate three-string junctions.  Thus, except for exceptional
states (represented by a single $n$-string junction in the string
picture), we can generically only tune the vacuum close to one CMS.
Thus our approximate boundary conditions will generically not be valid
for the whole low energy solution since the other junctions need not be
close to their CMS. 

However, having deduced the 3-string junction picture of the two source
states allows us to bootstrap our way to the string web picture of a
general $n$-source state.  To do this, group the $n$ sources in pairs
(in all possible ways).  Then solve for all 3-string junction states for
each of those pairs of charges to find the manifold of possible ``vacua"
where such 3-string junctions could originate.  We can then use these
manifolds of 3-string junction vacua as the values for a new set of
fuzzy ball boundary conditions, and thus determine the existence of a
new 3-string junction connecting pairs of the previously calculated
3-string junctions, and so on, thereby building up many-branched trees
of 3-string junctions.  This procedure thus constructs a picture of the
general BPS state as a string web on the moduli space of \Nf\ theories. 

\subsection{Projecting BPS prongs to string webs}\label{s4p2}

\EPSFIGURE{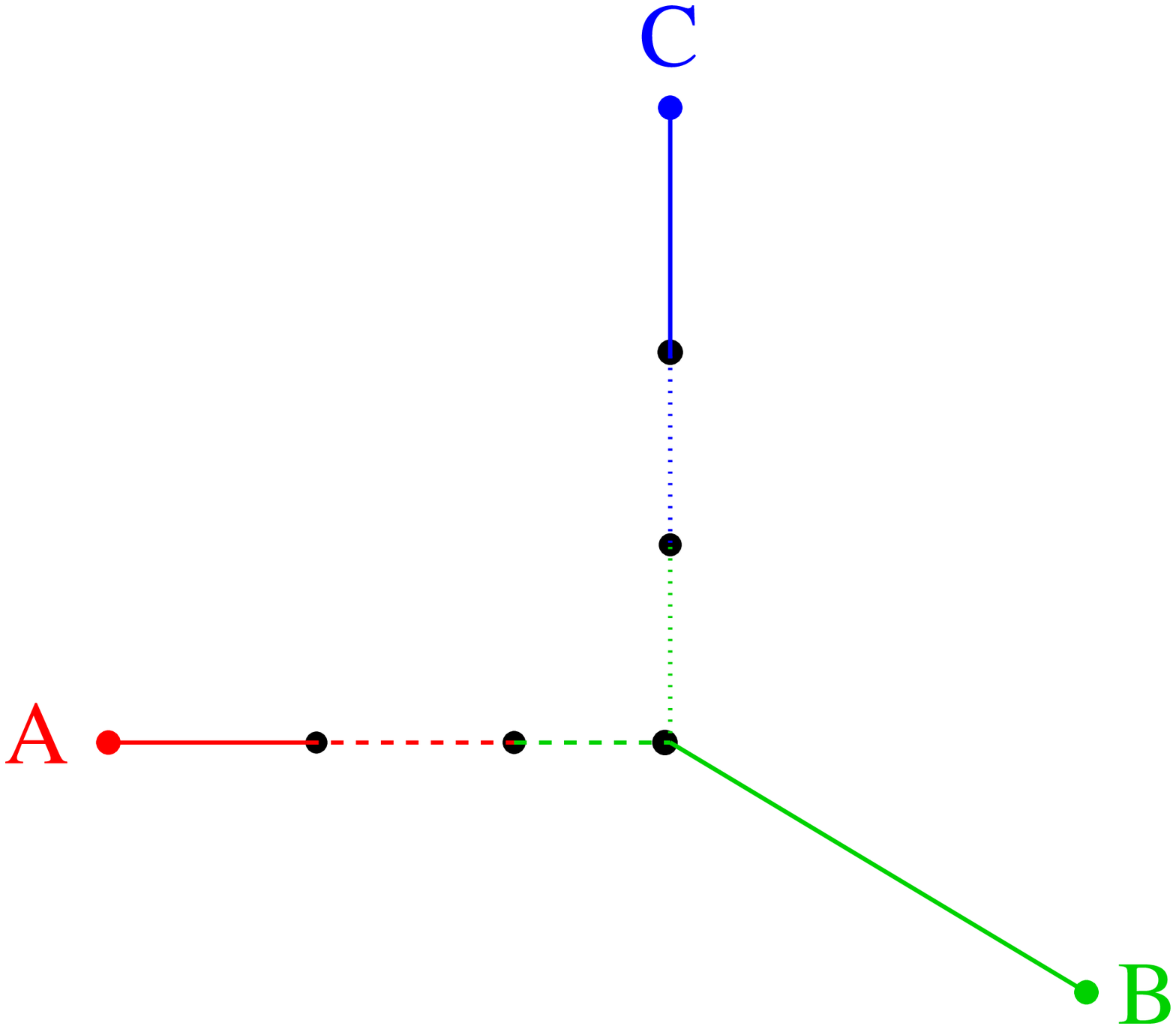,width=12em}{Image of a 3-string junction on moduli
space in a transverse plane to three D3-branes in IIB string theory.
The images of the $X^0\to\mbox{CMS}$ segment are solid lines, while
those of $X^1\to\mbox{CMS}$ are dashed, and those of $X^2\to\mbox{CMS}$
are dotted.  The segments traced out by the $A$ brane are shown in red,
the $B$ brane in green, and the $C$ brane in blue.\label{fig7}}

So far we have developed a picture of \Nf\ BPS states as string webs on
the moduli space of vacua.  However, the actual string web construction
of BPS states in string theory does {\em not} in general involve strings
on the moduli space of the theory.  For example, the $\SU(N)$ \Nf\
theory can be realized as open string and D1-brane degrees of freedom
propagating on a stack of parallel D3-branes in IIB string theory.  BPS
states are then represented as webs of these $(p,q)$ strings ending on
the D3-branes and extending in the transverse $\R^6$ spatial dimensions.
This $\R^6$ is not the moduli space of the \Nf\ theory, which is a
$6N$-dimensional space ${\cal M}=\R^{6N}/S_N$. 

However, there is a simple map relating points in the full moduli space
to arrangements of branes in the transverse $\R^6$.  We will now show
how this map can be used to map our string webs on the moduli space of
\Nf\ theories to the string webs in the $\R^6$ transverse to D3-branes
in IIB theory.  First, just as for our low energy solutions we could
restrict ourselves to a $\C^N/S_N$ subspace of $\cal M$, the condition
that 1/4 of the supersymmetry be left unbroken in the IIB string
construction allows one to confine oneself to string webs living in a
one-complex-dimensional subspace $\C$ of the transverse $\R^6$ space.
A general point $X^0$ in $\C^N/S_N$ is given as a complex vector with
$N$ unordered components, \eg\ $X^0=(a,b,c)$ in the case of $N=3$.  This
is mapped to a configuration of $N$ D3-branes at positions in the
complex transverse plane corresponding to the complex coordinates, \eg\
for $N=3$, $X^0$ corresponds to three D3-branes at positions $a$, $b$,
and $c$ in $\C$.  (Our freedom to set $a+b+c=0$ upon decoupling the
diagonal $\U(1)$ gauge factor corresponds in the brane picture to the
freedom to shift the center of mass position of the D3-branes to the
origin.)

This map can also be used to map string webs on the moduli space to
paths on the transverse $\C$.  As each point in moduli space translates
to an arrangement of D3-branes in $\C$, the image of a path in moduli
space will be a set of $N$ paths in $\C$ of each D3-brane.  Thus, in
the case of our 3-string junction (\ref{nfstringweb}) on moduli space,
each segment will translate into three segments in $\C$ traced out by
the corresponding D3-branes.  Since there are a total of three straight
segments on the moduli space, generically the image in $\C$ will consist
of nine segments.  We will now show that our BPS solution is just such
that these nine segments line up to form a 3-string junction on $\C$,
thus recovering the usual IIB string web. 

To see this, simply trace out the path of each D3-brane.  Starting at
the vacuum, $X^0=(a,b,c)$, we label the three D3-branes corresponding to
the points $a$, $b$, and $c\in\C$, as ``$A$", ``$B$", and ``$C$",
respectively.  Let us first follow the image of the $B$ D3-brane in
$\C$.  Since along the $X^0\to\mbox{CMS}$ segment the second coordinate
on the moduli space traverses $b\to c_1\bee_1+a_2\bee_2$, so therefore
does the $B$ brane in $\C$.  We can similarly trace out its image path
in $\C$ of the other segments.  Doing this for the $A$ and $C$ branes as
well we get the nine segments
\be
\begin{array}{|c||l|l|l|}
\hline
 & X^0\to\mbox{CMS} & \mbox{CMS}\to X^1 & \mbox{CMS}\to X^2 \\
\hline
\hline
A & 
a \to -2c_1\bee_1+ a_2\bee_2 & 
-2c_1\bee_1+ a_2\bee_2 \to -{1\over2}c_1\bee_1+ a_2\bee_2 &
-2c_1\bee_1+ a_2\bee_2 \to -2c_1\bee_1+         a_2\bee_2 \\
B & 
b \to + c_1\bee_1+ a_2\bee_2 &
+ c_1\bee_1+ a_2\bee_2 \to -{1\over2}c_1\bee_1+ a_2\bee_2 &
+ c_1\bee_1+ a_2\bee_2 \to + c_1\bee_1-{1\over2}a_2\bee_2 \\
C & 
c \to + c_1\bee_1-2a_2\bee_2 &
+ c_1\bee_1-2a_2\bee_2 \to +         c_1\bee_1-2a_2\bee_2 &
+ c_1\bee_1-2a_2\bee_2 \to + c_1\bee_1-{1\over2}a_2\bee_2 \\
\hline
\end{array}
\ee
Notice first that the $A$: $\mbox{CMS}\to X^2$ segment and the $C$:
$\mbox{CMS}\to X^1$ segment are both just points in $\C$.  Second, notice
that the $A$: $X^0\to\mbox{CMS}$ segment plus the $A$: $\mbox{CMS}\to X^1$
segment plus the $B$: $X^1\to\mbox{CMS}$ segment form a straight line $a
\to c_1\bee_1+a_2\bee_2$.  (Recall from (\ref{order}) and the fact that
$a_1+b_1+c_1=0$ that $c_1>-{1\over2}c_1> -2c_1>a_1$, so the segments are
traced out consecutively.) Similarly, the $C$: $X^0\to\mbox{CMS}$ segment
plus the $C$: $\mbox{CMS}\to X^2$ segment plus the $B$: $X^2\to\mbox{CMS}$
segment form a straight line $c \to c_1\bee_1+a_2\bee_2$.  Finally, the
$B$: $X^0\to\mbox{CMS}$ segment forms a third line to the same point.
Thus the image in $\C$ of the 3-string junction on moduli space is again
a three string junction leading from the vacuum positions $a$, $b$, $c$,
of the D3-branes to a common central point; see figure~\ref{fig7}. 

We have thus recovered the IIB string web picture of 1/4 BPS states
of the \Nf\ $\SU(N)$ superYang-Mills theory.

\section{BPS states in \Nt\ theories}\label{s5}

A similar picture of 1/2 BPS states as string webs on the Coulomb branch
of \Nt\ theories can be derived in close analogy to our discussion of
\Nf\ theories in the last section.  The generic point on the Coulomb
branch of an \Nt\ gauge theory with gauge group $G$ is described by a
low energy $\U(1)^N$ \Nt\ theory where $N=\mbox{rank}\,G$.  The theory
may have $N_f$ ``matter'' hypermultiplets in various representations;
they are massive at the generic Coulomb branch vacuum.  Each $\U(1)$
\Nt\ multiplet has one complex scalar field, $\phi_r$ $r=1,\ldots,N$,
whose vevs parameterize the Coulomb branch.  The other bosonic massless
fields are the electric and magnetic fields $\vE_r$ and $\vB_r$ for each
$\U(1)$ factor.  We are interested in static low energy solutions
carrying electric and magnetic charges $(n_E^r, n_{Br})$ with respect to
these fields.  We normalize the fields so that the low energy effective
action for the bosonic fields is\footnote{We have chosen the normalization
of the scalars $\phi_r$ to differ from their canonical normalization
by a factor of $\sqrt{2}$ to simplify later formulas.}
\be\label{ntact}
S= \int d^4x\left\{ {\Ima\tau^{rs}\over16\pi} F_{r\,\mu\nu}
F^{\mu\nu}_s + {\Rea\tau^{rs}\over32\pi} \epsilon_{\mu\nu\rho\sigma}
F_r^{\mu\nu}F_s^{\rho\sigma} + {\Ima\tau^{rs}\over16\pi}
\del_\mu \phi_r\del^\mu \phi_s^* \right\},
\ee
with an implicit sum over repeated indices, and where
$\vE_r$ and $\vB_r$ are related to $F^{\mu\nu}_r$ in the usual way,
and
\be
\tau^{rs}(\phi) \equiv {\vartheta^{rs}(\phi) \over2\pi}
+ i{4\pi\over g^2_{rs}(\phi)}
\ee
is the matrix of effective $\U(1)$ couplings and theta angles
which depend holomorphically on the $\phi_r$.  Thus the
metric on the Coulomb branch is given by the line element
\be
ds^2 = \Ima \tau^{rs}(\phi) \, d\phi_r d\phi_s^* .
\ee
Note that $\tau^{rs}$ is a symmetric matrix by definition, and that
$\Ima\tau^{rs}$ is positive definite by unitarity.
The rigid special Kahler geometry of the \Nt\ Coulomb branch
implies that
\be
\tau^{rs}(\phi) = {\del^2{\cal F}(\phi) \over \del\phi_r\del\phi_s}
\ee
for some prepotential $\cal F$.  

It is useful to introduce the ``dual scalars" 
\be\label{ntphidual}
\phi_D^r \equiv {\del{\cal F}\over\del\phi_r},
\ee
the complex field strength
\be
\vF_r \equiv \vB_r + i \vE_r
\ee
which appears in the same supermultiplet as $\phi_r$,
as well as its dual
\be\label{ntfdual}
\vF_D^r \equiv \tau^{rs}\vF_s ,
\ee
which is in the same multiplet as $\phi_D^r$.

The Bianchi identities as well as the static equations of motion in
vacuum (without sources) for the electric and magnetic fields that
follow from (\ref{ntact}) are
\bea\label{ntmaxwell}
0 &=& \vn\cdot\Rea\vF_D^r,
\nonumber\\
0 &=& \vn\cdot\Rea\vF_r,
\nonumber\\
0 &=& \vn\times\Ima\vF_D^r,
\nonumber\\
0 &=& \vn\times\Ima\vF_r,
\eea
implying, together with the Dirac quantization condition, that the
electric and magnetic charges are quantized as
\bea\label{ntgauss}
\oint_{S^2} \Rea\vF_D^r\cdot d\vec a &=& -4\pi n_E^r ,\nonumber\\
\oint_{S^2} \Rea\vF_r  \cdot d\vec a &=& +4\pi n_{Br},
\eea
for some integers $n_E^r$ and $n_{Br}$.  (The magnetic field
contribution in the first line includes the effect of the theta angle on
electric charge quantization.)

Just as in our discussion in section~\ref{s3} the scalar fields should
obey fuzzy ball boundary conditions at the charge sources, \ie\ their
values in the vicinity of a charge source approaches those of the
coordinates of a singularity on the Coulomb branch where a state of that
charge becomes massless.  The generic singularities on an \Nt\ Coulomb
branch are complex codimension one submanifolds, which we denote ${\cal
S}^{(i)}$, at which a state of given charges $(n_E^{(i)r},
n^{(i)}_{Br})$ becomes massless.  In a solution with $M$ charge sources,
we denote by $(n_E^{(i)r}, n^{(i)}_{Br})$ the charges of each source,
$i=1,\ldots,M$, and by $(n_E^{(0)r}, n^{(0)}_{Br})$ the total charge, so
that
\be\label{ntchrgcons}
(n_E^{(0)r}, n^{(0)}_{Br}) + 
\sum_{i=1}^M (n_E^{(i)r}, n^{(i)}_{Br}) = 0 .
\ee

The relation between the charges of the states which become
massless at ${\cal S}^{(i)}$ and the coordinate values of the
$\phi_r$ on ${\cal S}^{(i)}$ follows from the monodromies that
the $\phi_r$ experience upon traversing a path encircling 
${\cal S}^{(i)}$.  To see this, recall \cite{sw9408} that
the monodromy in the $\U(1)^N$ effective theory is an
element of $\Sp(2n,\Z)\ltimes \Z^{N_f}$ which acts on the
scalar fields and their duals, as well as the electric, magnetic,
and hypermultiplet number charges.  The latter are the integer charges
$n_Q^p$, $p=1,\ldots,N_f$, under the $\U(1)^{N_f}$ global flavor
symmetry group.  If the (bare) masses of the hypermultiplets are
denoted by $m_p$, then the monodromy acts on the scalar fields
as
\be\label{ntmonod1}
\pmatrix{\phi_D^r \cr\phi_r\cr} \to
{\bf S} \pmatrix{\phi_D^r \cr\phi_r\cr} + 
{\bf T}\pmatrix{m_1\cr \vdots\cr m_{N_f}\cr},
\ee
where
${\bf S}\in\Sp(2N,\Z)$ and ${\bf T}$ is a $2N\times N_f$ integer
matrix.  The monodromy matrices around a submanifold ${\cal S}$
where a state of charges $(n_E^r, n_{Br}, n_Q^p)$ is massless is
then \cite{aps9505}
\bea\label{ntmonod2}
{\bf S} &=& \pmatrix{\delta^r_s + n_E^r n_{Bs} & n_E^r n_E^s\cr
-n_{Br} n_{Bs} & \delta^s_r - n_{Br} n_E^s \cr} ,\nonumber\\
{\bf T} &=& \pmatrix{n_Q^p n_E^r \cr - n_Q^p n_{Br}\cr} .
\eea
The coordinates $\phi_r$ or $\phi_D^r$ of the singularity ${\cal S}$
are characterized by the condition that they be invariant
under this monodromy.  This condition together with (\ref{ntmonod1}) 
and (\ref{ntmonod2}) then imply that 
\be\label{ntSeqn}
n_E^r \phi_r + n_{Br} \phi_D^r + n_Q^p m_p = 0.
\ee
This one complex equation (locally) determines the complex-codimension
one singular manifold ${\cal S}$, and will play an important role
in what follows.

(There is a global issue concerning the identification of the charges of
the state which becomes massless.  As we have discussed extensively in
earlier sections, the static low energy BPS solutions can be thought of
as a map from three-space into the Coulomb branch interpolating between
the vacuum (at $\vx=\infty$) and various singularities ${\cal S}^{(i)}$
(at points $\vx=\vx_i$).  Since $\tau^{rs}$ can undergo $\Sp(2N,\Z)$
monodromies upon traversing paths which loop around singularities in the
Coulomb branch, the charges of the state which becomes massless at
${\cal S}^{(i)}$ must be redefined by the monodromy corresponding to the
path on the Coulomb branch traced out by the low energy solution as it
interpolates between $\vx=\infty$ and $\vx_i$.  Thus the integers
$(n_E^{(i)r}, n^{(i)}_{Br})$ will refer to the charges referred back
in this way to $\vx=\infty$.)

With this description of the moduli space and its singularities in hand,
we are ready to solve for static low energy field configurations in a
given charge sector subject to our fuzzy ball boundary conditions.  The
mass of our configuration is computed by integrating the field energy
density (with an implicit sum over repeated indices)
\be
M = {1\over8\pi}\int d^3\vx\, \Ima(\tau^{rs})(\vF_r\cdot\vF^*_s  
+ \vn\phi_r\cdot \vn\phi^*_s ).
\ee
This can be rewritten \cite{cru9612} by the usual trick of completing
the square with surface terms:
\bea\label{ntdecomp}
M &=& {1\over8\pi}\int d^3\vx\, \Ima(\tau^{rs}) 
(\vF_r + e^{i\alpha} \vn \phi_r) \cdot
(\vF_s^* + e^{-i\alpha} \vn\phi^*_s) \nonumber\\
&& \ {}- {1\over4\pi}\int d^3\vx\, \Ima(\tau^{rs}) \left\{\vB_r
\cdot \Rea(e^{i\alpha} \vn\phi_s) +  \vE_r
\cdot \Ima(e^{i\alpha} \vn\phi_s)\right\} \nonumber\\
&\ge& -{1\over4\pi} \int d^3\vx \left\{ \Rea(\vF_r) \cdot
\Ima (e^{i\alpha} \vn\phi_D^r) - \Rea(\vF_D^r) \cdot
\Ima (e^{i\alpha} \vn\phi_r) \right\} \nonumber\\
&=& \Ima \sum_{I=0}^M e^{i\alpha} \left\{ n_{Br}^{(I)} \phi_D^{(I)r}
+ n_E^{(I)r} \phi_r^{(I)} \right\} .
\eea
Here $\alpha$ is a constant phase to be determined below.
The third line comes from integrating by parts and using
(\ref{ntgauss}).
The inequality arises because the first term in the first line
is positive, so that the inequality is saturated only if
the BPS equations
\be\label{ntBPS}
\vF_r + e^{i\alpha} \vn \phi_r = 0 
\ee
are satisfied.  It is useful to note, using (\ref{ntphidual}) and 
(\ref{ntfdual}) that this can be rewritten as
\be\label{ntBPSdual}
\vF_D^r + e^{i\alpha} \vn \phi_D^r = 0 .
\ee
It follows from these equations and (\ref{ntmaxwell}) that
away from the sources
\be\label{ntharmonic}
\nabla^2 \Rea(e^{i\alpha}\phi_r) =
\nabla^2 \Rea(e^{i\alpha}\phi_D^r) = 0 .
\ee

The BPS bound arises from maximizing the right hand side of
(\ref{ntdecomp}).  Define
\be
Z^{(I)} \equiv n_{Br}^{(I)} \phi_D^{(I)r} + n_E^{(I)r} \phi_r^{(I)},
\ee
the central charge associated with the $I$th source, and denote by
\be
Z \equiv \sum_{I=0}^M Z^{(I)}
\ee
their sum.  Then
\be\label{ntmbps1}
M_{BPS} = \max_\alpha \Ima\{e^{i\alpha}Z\}=|Z|,
\ee
implying that
\be\label{ntalpha}
e^{i\alpha} = i Z^*/|Z|.
\ee

This expression for the BPS mass should coincide with the
usual \Nt\ BPS mass formula \cite{sw9408}
\be\label{ntswbps}
M_{BPS} = | n_E^{(0)r} \phi_r^{(0)} + n_{Br}^{(0)} \phi_D^{(0)r}
+ n_Q^{(0)p} m_p | ,
\ee
where $n_Q^{(0)p}$ are the total ``quark'' numbers of the
state in question.
For (\ref{ntswbps}) and (\ref{ntmbps1}) to be the same requires
that
\be\label{ntA}
n_Q^p m_p = \sum_{i=1}^M Z^{(i)}.
\ee
But this follows automatically from the condition (\ref{ntSeqn})
characterizing the singularities on the Coulomb branch, which
reads
\be\label{ntB}
Z^{(i)} = -n^{(i)p}_Q m_p
\ee
where $n^{(i)p}_Q$ are the ``quark'' quantum numbers of the $i$th
charge source.  Conservation of quark number implies
\be\label{ntnqcons}
\sum_{I=0}^M n_Q^{(I)p} = 0,
\ee
which, together with (\ref{ntB}) implies (\ref{ntA}).

Now, as described above, the singularities of the Coulomb branch are
whole complex codimension one submanifolds ${\cal S}^{(i)}$, not
isolated points, so the coordinates $\phi_r^{(i)}$ of the
singularities can vary continuously.  We have just demonstrated that
the BPS mass bound is independent of the precise values of the
$\phi_r^{(i)}$.  Therefore, unlike the \Nf\ case, no minimization of
the BPS bound over the ${\cal S}^{(i)}$ needs to be performed.  But,
as in the \Nf\ case, the extra conditions needed to fix the boundary
conditions (the $\phi^{(i)}$) come from demanding that solutions to
the BPS equations exist.

So, suppose we fix the $\phi^{(i)}$, and thus the boundary values of
the scalars.  We will now solve the low energy BPS equations
(\ref{ntharmonic}) subject to the charge and fuzzy ball boundary
conditions.  In doing so we will determine necessary conditions for a
solution to exist given these boundary conditions.  As in the \Nf\
case it is sufficient to solve the equations with just one or two
charge sources, as solutions with more sources can be built up from
these simpler ones.  Also just as in the \Nf\ case we will show below
that the condition that these solutions exist can be phrased as the
conditions that a certain matrix $\alpha_{ij}$ be real, symmetric, and
have only positive entries.  The reality and symmetry conditions will
provide the extra conditions needed to determine the $\phi_r^{(i)}$
for one or two charge sources.  Finally, the positivity condition is
satisfied only on one side of the CMS, and so it is this condition
which determines the stability of BPS states.

The harmonic BPS equations (\ref{ntharmonic}) can be solved as in 
the toy model of section~\ref{s3}
by a superposition of single source solutions:
\bea\label{ntsoln3}
\Rea\left(e^{i\alpha}\phi_r(\vx)\right) &=& 
\sum_{i=1}^M {n_{Br}^{(i)} \over |\vx - \vx_i|} + 
\Rea\left(e^{i\alpha}\phi^{(0)}_r(\vx)\right),\nonumber\\
\Rea\left(e^{i\alpha}\phi^r_D(\vx)\right) &=& 
-\sum_{i=1}^M {n_E^{(i)r} \over |\vx - \vx_i|} +
\Rea\left(e^{i\alpha}\phi^{(0)r}_D(\vx)\right) .
\eea
The numerators on the right hand side are determined by (\ref{ntBPS})
and (\ref{ntgauss}).  Now, the fuzzy ball boundary conditions are that
$\phi_r$ goes through a small ball around $\phi_r^{(i)}$ as $\vx\to\vx_i$.
In this limit we have
\bea
\lim_{\vx\to\vx_i} \Rea\left( e^{i\alpha}\phi_r(\vx)\right)&\simeq& 
{n_{Br}^{(i)} \over \epsilon_i} + 
\sum_{j \neq i}^M {n_{Br}^{(j)}\over r_{ij}} +
\Rea\left( e^{i\alpha}\phi_r^{(0)}\right), \nonumber\\
\lim_{\vx\to\vx_i} \Rea\left( e^{i\alpha}\phi_D^r(\vx)\right)&\simeq& 
-{n_E^{(i)r} \over \epsilon_i} - 
\sum_{j \neq i}^M {n_E^{(j)r}\over r_{ij}} +
\Rea\left( e^{i\alpha}\phi_D^{(0)r}\right),
\eea
where 
\be\label{ntdefepr}
\epsilon_i \equiv |\vx - \vx_i|, 
\quad \mbox{and}\quad
r_{ij} \equiv |\vx_i - \vx_j|.
\ee
Thus as $\vx\to\vx_i$ our solutions go to infinity asymptoting a line
in the $\Rea e^{i\alpha}\phi_r$--$\Rea e^{i\alpha}\phi_D^r$ plane along 
the $(n_{Br}^{(i)}, n_E^{(i)r})$ direction and with some intercept.
So a necessary condition for the fuzzy ball boundary
conditions to be satisfied is that the approximate boundary value
$\phi^{(i)}_r$ at the $i$th source lies on this asymptote:
\bea\label{ntdefalf}
\Rea\left( e^{i\alpha}\phi_r^{(i)}\right) &=& \sum_j \alpha_{ij}  
n_{Br}^{(j)} + \Rea\left( e^{i\alpha}\phi_r^{(0)}\right), \nonumber\\
\Rea\left( e^{i\alpha}\phi_D^{(i)r}\right)&=&-\sum_j \alpha_{ij} 
n_E^{(j)r} + \Rea\left( e^{i\alpha}\phi_D^{(0)r}\right),
\eea
for some $\alpha_{ij}$ a real symmetric matrix of positive numbers,
\be\label{ntdefalf2}
\alpha_{ij} = \alpha_{ji} >0,
\ee
which follows because $\alpha_{ij}=1/r_{ij}$ for $i\neq j$ are the
spatial source separations, while $\alpha_{ii}=1/\epsilon_i$ is
the spatial cutoff around the $i$th source.

The conditions (\ref{ntdefalf}) subject to (\ref{ntdefalf2}) are $2MN$
real equations for $2MN-M(3-M)/2$ real unknowns.  (There are
$M(M+1)/2$ $\alpha_{ij}$'s and $2N-2$ real components of
$\phi^{(i)}_r$ in each of the $M$ ${\cal S}^{(i)}$ singular
submanifolds.)  In particular, for the one source $M=1$ (``spike'')
and two source $M=2$ (``prong'') solutions in which we are interested,
we have one more equation than unknown, which would imply that
generically there are no solutions to the BPS equations and our
boundary conditions.  However, there is one combination of these
equations which is automatically satisfied by virtue of the condition
(\ref{ntSeqn}) defining the singular manifolds ${\cal S}^{(i)}$.  In
particular, the sum over $i$ and $r$ of $n_E^{(i)r}$ times the first
equation in (\ref{ntdefalf}) plus that of $n_{Br}^{(i)}$ times the
second gives an identity using (\ref{ntSeqn}) as well as charge
conservation (\ref{ntchrgcons}), (\ref{ntnqcons}), and the definition
(\ref{ntalpha}) of the phase $e^{i\alpha}$.  Therefore there is
generically a unique spike or prong solution to the BPS equations and
our boundary conditions which uniquely fixes the boundary values of the
$\phi^{(i)}_r$.  Just as in the \Nf\ case the positivity condition on
the $\alpha_{ij}$ will prevent some solutions from existing; places
where a solution ceases to exist because one of the $\alpha_{ij}$
changes sign will correspond to the decay of a state across a CMS.

It now only remains to show that the projection of these spike and
prong solutions onto the Coulomb branch leads to a string junction
picture of the states, as least in the limit as the vacuum approaches
the appropriate CMS.  But this follows with almost no work from what
we have done so far.

In particular, a one source (spike) solution is rotationally
invariant about its spatial center $\vx_1$; therefore the projection
of the solution (\ref{ntsoln3}) on the Coulomb branch is just
a one real dimensional path leading from the vacuum to the singular 
submanifold.  In the case of a one complex dimensional
Coulomb branch ($N=1$), taking the sum of $n_E$ times the first equation
in (\ref{ntsoln3}) plus $n_B$ times the second gives the condition
(satisfied everywhere along the solution)
\be
\Rea\left( e^{i\alpha} [n_E \phi + n_B \phi_D + n_Q\cdot m]\right)=0.
\ee
But this is precisely the condition determining the path of an
$(n_E,n_B)$ string stretched on the $\SU(2)$ Coulomb branch found in
its F-theory representation \cite{f9701,mns9803,bf9806}, and implies
in particular that they lie along geodesics in the Coulomb branch
metric.  We thus recover the string picture for the spike solutions in
the case of a one complex dimensional Coulomb branch.  Furthermore, in
the cases where F theory describes a higher-dimensional Coulomb branch
(\ie\ with multiple D3-brane probes of a 7-brane background
corresponding to the $\Sp(2N)$ theory with four fundamental and one
massless antisymmetric hypermultiplet \cite{asty9611,dls9612} and the
higher rank generalizations of the exceptional theories with $E_n$
global symmetry groups \cite{fs9805,afm9806}), the Coulomb branch is
just the tensor product of one complex dimensional Coulomb branches
(modulo permutations) just as with the \Nf\ moduli space.  In these
cases the F theory string picture is found by a similar mapping of the
``string'' found by projecting the spike solution onto the
multi-dimensional Coulomb branch onto the one dimensional space.
Note, however, that our string picture persists in theories with
higher-rank Coulomb branches even when they do {\em not} have a
direct product geometry.

Finally, it remains to show that the projection of the brane prong 
(two source) solutions onto the Coulomb branch recovers the string
junction picture of these states in the vicinity of the CMS.  We will
illustrate this in a simple example below, but the main point follows
from our general discussion so far:  the diverging spatial separation
of the charge centers as we approach the CMS implies that any prong
state will more and more accurately approach the sum of two spike
solutions, whose Coulomb branch projections we have just seen are 
$(n_E,n_B)$ strings lying along geodesics.  Furthermore, in the
limit as the vacuum approaches the CMS, the analysis of the solution 
in a neighborhood of the vacuum point will reproduce precisely the
string junction conditions with the junction lying on the CMS.
This follows because in this limit we can ignore the curvature of the
Coulomb branch metric in a small enough neighborhood of the vacuum
point, thus taking $\tau^{rs}$ to be locally constant,
so that $\phi_D^r \sim \tau^{rs}\phi_s$.  Then the \Nt\ solutions
(\ref{ntsoln3}) become formally the same as the \Nf\ solutions
(\ref{nfsoln3}), and so inherit their local properties.

This can also be checked explicitly using some detailed properties
of a given Coulomb branch geometry.  We will illustrate this in the
simplest example, namely the decay of the $W$ boson in $\SU(2)$
\Nt\ superYang-Mills across its CMS.  We follow the conventions
of \cite{bf9802} (our $\phi$ and $\phi_D$ correspond to their
$a$ and $a_D$).  The massive $W$ boson is a BPS state with charges
$(n_E^{(0)},n_B^{(0)})=(2,0)$ and so in a vacuum with coordinates
$\phi$ or $\phi_D$, has central charge
\be
Z = 2 \phi,
\ee
implying that 
\be
e^{i\alpha} = i {\phi^*\over |\phi|} .
\ee
(We have dropped the $(0)$ superscripts on the $\phi$ and $\phi_D$
coordinates for the vacuum.)  The CMS is given by a curve satisfying
the real condition
\be
\Ima\left({\phi_D\over\phi}\right) = 0 .
\ee
The Coulomb branch has two singularities, a ``magnetic'' one with 
coordinates
\be
\phi^{(1)}={2\over\pi}, \qquad \phi_D^{(1)}= 0
\ee
where a state of charge
\be
(n_E^{(1)},n_B^{(1)})=(0,1)
\ee
becomes massless; and a ``dyonic'' one with coordinates
\be
\phi^{(2)}={2i\over\pi}, \qquad \phi_D^{(2)}= -{4i\over\pi}
\ee
where a state of charge
\be
(n_E^{(2)},n_B^{(2)})=(-2,-1)
\ee
becomes massless.  Plugging all this into the conditions (\ref{ntdefalf})
gives three independent equations for the $\alpha_{ij}$ which can be solved
to give
\bea
\alpha_{12} &=& {|\phi|\over2} \Ima\left({\phi_D\over\phi}\right),
\nonumber\\
\alpha_{11} &=& {|\phi|\over2} \Ima\left({\phi_D\over\phi}\right)
+ {2\over\pi} {\Ima\phi\over|\phi|} ,\nonumber\\
\alpha_{22} &=& {|\phi|\over2} \Ima\left({\phi_D\over\phi}\right)
+ {2\over\pi} {\Rea\phi\over|\phi|} .
\eea
The positivity condition for $\alpha_{12}$ then implies
\be
\Ima\left({\phi_D\over\phi}\right) > 0,
\ee
which is precisely the condition to be outside the CMS.
Furthermore, as we approach the CMS, $\alpha_{12}\to0$,
which, since $\alpha_{12} = 1/r_{12}$, implies that
the charge sources are becoming infinitely spatially
separated in the limit.

We have thus seen how to recover the string web picture
of \Nt\ BPS states from the low energy $\U(1)^N$ effective
action on the Coulomb branch.  We have checked that it matches
precisely with the string webs found in the F theory realization
of certain \Nt\ theories as D3-brane probes of 7-brane
backgrounds.  But we emphasize that our string web picture
has much greater generality:  we have shown that BPS states
can be represented by string webs on the Coulomb branch of
an {\em arbitrary} \Nt\ theory, irrespective of whether
a string construction of that theory is known.

\section{Open questions and future directions}\label{s6}

In this paper we have shown that a string junction picture of
decaying BPS states follows directly from the low energy effective
theory in the vicinity of a CMS and relies on approximate boundary
conditions which become exact in the limit of approaching the CMS.
Furthermore this construction works for arbitrary field theory
data (gauge group, matter representations, couplings and masses).

There are, however, a number of questions arising in this framework
which we have not addressed.  The most pressing of these is the
``s-rule'' \cite{hw9611,mns9803,dhiz9805,bf9806} in \Nt\ theories, which is not
apparent in our solutions.  The s-rule is a selection rule on string
junction configurations in 7-brane backgrounds which rules out some
string junctions as not being supersymmetric even though they satisfy
the charge conservation and tension-balancing constraints.  Our
generalized string webs reproduce these constraints but not the s-rule
constraints.  The s-rule presumably arises from the low energy point
of view as an extra condition on our \Nt\ solutions (\ref{ntsoln3}) to
be BPS.  In fact, such an extra condition may already be present in
our discussion of the \Nt\ BPS states, for we really solved only the
harmonic equations (\ref{ntharmonic}) following from the BPS equations
(\ref{ntBPS}) and the Gauss constraints (\ref{ntgauss}).  To rederive
the BPS equations from the harmonic equations requires an extra
condition, namely that $\phi_r$ and $\phi_D^r$ be related by the
special geometry relation (\ref{ntphidual}).\footnote{We thank
A. Shapere for discussions on this point.}  Perhaps this extra
condition gives rise to the s-rule for our low energy configurations.
A low energy understanding of the s-rule is also needed to examine
proposed stable non-BPS states in \Nt\ theories \cite{b9811}.

There are also a number of other issues which can be examined in our
framework.  One is to use our string web picture to compute the
spectrum of stable BPS states at strong coupling and conformal points
in higher rank \Nt\ theories; in some cases such spectra have been
proposed using other methods \cite{gh9906,sv9910,l0006}.  Another open
question which should be addressable in our framework is the question
of the multiplicity of stable BPS states, for which only partial
information is known from semiclassical techniques
\cite{h9510,bk9804,k9812,sy0005}.

It is an open question whether the string T-dual version of the string
web picture of BPS states---namely BPS states as curves on the
Seiberg-Witten Riemann surface with a certain metric
\cite{klmvw9604,w9703,hy9707,m9708,gh9906,sv9910,l0006}---has a low
energy field theory explanation, and if so, what its relation is to
the string web picture we have derived here.

Finally, our string web picture can be generalized in a number of
directions.  Our basic physical picture for the existence of a string
web picture of BPS states near a CMS was very general and did not
depend on the dimension of space-time.  It would be interesting to
apply these arguments to CMS in two dimensional \cite{cv9211}, three
dimensional \cite{hw9611}, and five dimensional \cite{ah9704,ahk9710}
theories, especially as string constructions in the latter two cases
already provide a string web picture of BPS states in certain cases.
Another interesting generalization is to effective gravitational
theories \cite{d0005} and the associated question of the connection to
formulations of BPS stability in string theory compactifications
\cite{km9908,dfr0002}.

\acknowledgments It is a pleasure to thank A.~Buchel, J.~Hein,
R.~Maimon, J.~Maldacena, M.~Moriconi, S.~Pelland, M.~Rangamani,
V.~Sahakian and A.~Shapere for helpful comments and discussions.  This
work was supported in part by NSF grant PHY95-13717.

\appendix
\section{Appendix}\label{sA}

This appendix solves the conditions coming from the extremization of the
BPS bound, (\ref{nfbps3}), and the reality and symmetry of the
$\alpha_{ij}$ in (\ref{defalf}) in the case of two charge sources.  The
setup is as described in section~\ref{s4p1prong}, so points on the
$\SU(3)$ moduli space are parameterized by complex 3-vectors $(a,b,c)$
satisfying $a + b + c = 0$.  The low energy $\U(1)^2$ charges are also
described by complex 3-vectors whose components sum to zero.  Vectors of
this form can be transformed into complex 2-component vectors by a
convenient unitary transformation
\be
{1\over\sqrt3} 
\pmatrix{\sqrt2 & -{1\over\sqrt2} & -{1\over\sqrt2} \cr
0 & \sqrt{3\over2} & -\sqrt{3\over2} \cr 1&1&1\cr}
\pmatrix{a\cr b\cr c\cr}
= {1\over\sqrt2} \pmatrix{\sqrt3 a \cr b-c \cr 0\cr}
\ee
which rotates the third component to zero.
Then the charges become
\be
Q_1 = {1\over\sqrt2} q_1 \pmatrix{\sqrt3\cr -1\cr}, \qquad
Q_2 = {1\over\sqrt2} q_2 \pmatrix{0\cr 2\cr},
\ee
the manifolds of singularities become
\be
X^1 ={1\over\sqrt2} x \pmatrix{\sqrt3\cr 3\cr},\qquad
X^2 ={1\over\sqrt2} y \pmatrix{-2\sqrt3\cr 0\cr},
\ee
and the vacuum becomes
\be
X^0 = {1\over\sqrt2} \pmatrix{\sqrt3 a \cr b-c \cr}.
\ee
The vectors pointing from the vacuum to the singularities
are thus
\be
\xi^1 ={1\over\sqrt2} \pmatrix{\sqrt3x-\sqrt3a\cr 3x+c-b\cr},\qquad
\xi^2 ={1\over\sqrt2} \pmatrix{-2\sqrt3y-\sqrt3a\cr c-b\cr}.
\ee

The BPS bound to be extremized is
\bea\label{appBPS}
M &=& \Rea\{\xi^1\cdot A\cdot Q_1^* + \xi^2\cdot A\cdot Q_2^*\}\\
&=& {1\over2} \Rea\Bigl\{ 
q_1^* \biggl[ x (3A_{11} -\sqrt3 A_{12} +3\sqrt3 A_{21} - 3A_{22}) 
- 3aA_{11} + \sqrt3 aA_{12} \nonumber\\
&& \qquad\qquad +(c-b) (\sqrt3 A_{21} - A_{22}) \biggr]
-q_2^* \biggl[ 4\sqrt3 yA_{12} + 2\sqrt3 aA_{12} + 2(b-c)A_{22} \biggr] 
\Bigr\} \nonumber
\eea
where $A_{ij}$ is a general $\U(2)$ matrix.  Extremizing with respect to
$x$ and $y$ first implies that $A_{12}=A_{21}=0$ and $A_{11}= A_{22}=
e^{i\phi}$ with $\phi$ undetermined.  Maximizing with respect to $\phi$
gives the BPS mass as
\bea
M_{BPS} &=& \max_\phi \Rea \left\{ e^{i \phi} 
[ q^*_1 (b-a) + q^*_2 (c-b) ]\right\} \nonumber\\
&=& |q^*_1 (b-a) + q^*_2 (c-b)|.
\eea
The maximization in the first line implies that $\phi$
is the phase of $q_1(b^*-a^*)+q_2(c^*-b^*)$, or equivalently,
that
\be\label{appphi}
0 = \Ima \left\{ e^{i \phi} [ q^*_1 (b-a) + q^*_2 (c-b) ]\right\}.
\ee

To determine $x$ and $y$ we must perform the extremization with
respect to the $\U(2)$ matrix $A$ before doing the $x$, $y$ extremization.
Expand $A$ about its extremal value as
\bea
A = e^{i \phi}
\pmatrix{1 + iA   & iB - C \cr iB + C & 1 +iD \cr}
\eea
for small real $A$, $B$, $C$ and $D$.  Inserting this in the
BPS bound (\ref{appBPS}) and extremizing with respect to $A$, $B$,
$C$ and $D$ gives the four conditions
\bea\label{appcond1}
0 &=& \Ima \left\{ e^{i\phi} q^*_1 (x-a)\right\}, 
\nonumber\\
0 &=& \Ima \left\{ e^{i\phi} [q^*_1 (3x+c-b)+2q_2^*(b-c)] \right\}, 
\nonumber\\
0 &=& \Ima \left\{ e^{i\phi} [q^*_1 (x-b)-q_2^*(2y+a)] \right\}, 
\nonumber\\
0 &=& \Rea \left\{ e^{i\phi} [q^*_1 (2x+c)+q_2^*(2y+a)] \right\}.
\eea
Subtracting three times the first equation from the second
gives back the condition (\ref{appphi}) which determines $\phi$.
Thus a convenient set of independent conditions on $x$ and $y$
can be taken to be
\bea\label{appcond2}
0 &=& \Ima \left\{ e^{i\phi} q^*_1 (x-a)\right\}, 
\nonumber\\
0 &=& \Ima \left\{ e^{i\phi} q_2^* (y-c)\right\}, 
\nonumber\\
0 &=& \Rea \left\{ e^{i\phi} [q^*_1 (2x+c)+q_2^*(2y+a)] \right\},
\eea
where the second condition is a linear combination of the first
three equations of (\ref{appcond1}).  These are three real equations
for the two complex variables $x$ and $y$.

The extra condition needed to determine $x$ and $y$ completely
comes from the reality and symmetry of the $\alpha_{ij}$
coefficients defined in (\ref{defalf2}).  Inserting our two-component
vector values for the $\xi^i$ and $Q^i$ gives the equations
\bea
e^{i\phi} \pmatrix{\sqrt3 x-\sqrt3 a \cr 3x+c-b\cr}
&=& \alpha_{11} q_1 \pmatrix{\sqrt3\cr -1\cr} 
+ \alpha_{12} q_2 \pmatrix{0\cr 2\cr} ,\nonumber\\
e^{i\phi} \pmatrix{-2\sqrt3 y-\sqrt3 a \cr c-b\cr}
&=& \alpha_{21} q_1 \pmatrix{\sqrt3\cr -1\cr} 
+ \alpha_{22} q_2 \pmatrix{0\cr 2\cr} ,
\eea
which can be inverted to give
\bea
\alpha_{11} &=& e^{i\phi}(x-a)/q_1 , \nonumber\\
\alpha_{21} &=& e^{i\phi}(-2y-a)/q_1 , \nonumber\\
\alpha_{12} &=& e^{i\phi}(2x+c)/q_2 , \nonumber\\
\alpha_{22} &=& e^{i\phi}(-y+c)/q_2 .
\eea
The reality and symmetry conditions on the $\alpha_{ij}$ give
five more conditions: $\Ima \alpha_{ij}=0$ and 
$\Rea(\alpha_{12}-\alpha_{21})=0$.  In fact only one of
these conditions is independent of the three conditions
in (\ref{appcond2}) and the condition (\ref{appphi})
determining $\phi$.

To see this and to solve these conditions, it is convenient
to choose a non-orthogonal basis $\{\bee_1,\bee_2\}$ for the 
complex numbers defined by
\be
\bee_j \equiv e^{i(\theta_j-\phi)},
\ee
where $\theta_1$ and $\theta_2$ are defined to be the phases 
of $q_1$ and $q_2$:
\be
q_j = |q_j| e^{i\theta_j} .
\ee
We then write, for example, 
\be
a = a_1 \bee_1 + a_2 \bee_2
\ee
for unique real numbers $a_j$; define similarly the real numbers $b_j$,
$c_j$, $x_j$,and $y_j$, for $j=1,2$.

In this basis the conditions (\ref{appcond2}) become
\bea\label{appcond3}
0 &=& x_2-a_2,\nonumber\\
0 &=& y_1-c_1,
\eea
and
\be\label{appcond4}
0 = [2x_1+c_1+(2x_2+c_2)\cos(\theta_1-\theta_2)]|q_1|
+[2y_2+a_2+(2y_1+a_1)\cos(\theta_1-\theta_2)]|q_2|,
\ee
and
\be\label{phicond}
(a_2-b_2)|q_1| = (b_1-c_1)|q_2|,
\ee
where this last condition is the translation into this basis
of the condition (\ref{appphi}) determining $\phi$.
The conditions that $\Ima\alpha_{11}=\Ima\alpha_{22}=0$ are
easily seen to give the first two conditions in (\ref{appcond3})
again, and so are not independent.  The other two $\alpha_{ij}$
reality conditions give
\bea\label{appcond5}
\Ima \alpha_{12} = 0 &=& 2y_2+a_2,\nonumber\\
\Ima \alpha_{21} = 0 &=& 2x_1+c_1.
\eea
These two conditions, together with (\ref{appcond3}) and 
(\ref{phicond}), imply (\ref{appcond4}).  Thus
(\ref{appcond3}) and (\ref{appcond5}) are an 
independent set of conditions determining $x$ and $y$
to be
\bea
x &=& -{1\over2} c_1 \bee_1 + a_2 \bee_2,\nonumber\\
y &=& +c_1 \bee_1 -{1\over2} a_2\bee_2.
\eea

Finally, we must show that the symmetry condition on
the $\alpha_{ij}$, namely $\Rea(\alpha_{12}-\alpha_{21})=0$,
is not an independent condition.  This condition reads
in our basis
\be
0 = |q_1| [2x_2+c_2+(2x_1+c_1)\cos(\theta_1-\theta_2)]
+ |q_2| [2y_1+a_1+(2y_2+a_2)\cos(\theta_1-\theta_2)] ;
\ee
plugging in the values for $x$ and $y$ then gives
precisely (\ref{phicond}), which shows that it is
indeed not an independent condition.

\end{document}